\documentclass[%
 aip,
 amsmath,amssymb,
 reprint,%
]{revtex4-1}
\usepackage[normalem]{ulem} 
\usepackage{soul} 
\usepackage{graphicx}
\usepackage{dcolumn}
\usepackage{bm}
\usepackage{euscript,stmaryrd}
\usepackage{xcolor}
\usepackage[utf8]{inputenc}
\usepackage[T1]{fontenc}
\usepackage{mathptmx}
\usepackage{epsf,amsmath,amssymb,verbatim,color,multirow,pifont,mathrsfs,soul,amsthm, wasysym,bm}
\usepackage{graphicx, upgreek}
\usepackage[us,12hr]{datetime}
\usepackage{xcolor}
\usepackage{arydshln}
\usepackage{hyperref}
\usepackage{arydshln}
\usepackage{graphicx}
\usepackage{mathtools}

\theoremstyle{definition}
\newtheorem{myde}{Conjecture}
\newtheorem{mydef}[myde]{Condition}

\newcommand\redout{\bgroup\markoverwith
{\textcolor{red}{\rule[0.5ex]{2pt}{0.8pt}}}\ULon}

\newcommand\coolover[2]{\mathrlap{\smash{\overbrace{\phantom{%
    \begin{matrix} #2 \end{matrix}}}^{\mbox{$#1$}}}}#2}

\usepackage{hyperref}
\hypersetup{
    colorlinks,
    citecolor=blue,
    filecolor=blue,
    linkcolor=blue,
    urlcolor=black
}

\begin{document}

\preprint{AIP/123-QED}

\title[Attracting Poisson Chimeras in Two-population Networks]
{Attracting Poisson Chimeras in Two-population Networks}

\author{Seungjae Lee}
\email{seungjae.lee@tum.de}
\affiliation{Physik-Department, Technische Universit\"at M\"unchen, James-Franck-Stra\ss e 1, 85748 Garching, Germany}
\author{Katharina Krischer}%
\email{krischer@tum.de}
\affiliation{Physik-Department, Technische Universit\"at M\"unchen, James-Franck-Stra\ss e 1, 85748 Garching, Germany}

\date{\today}

\begin{abstract}
Chimera states, i.e., dynamical states composed of coexisting synchronous and asynchronous oscillations, have been reported to exist in diverse topologies of oscillators in simulations and experiments. Two-population networks with distinct intra - and inter-population coupling have served as simple model systems for chimera states since they possess an invariant synchronized manifold, in contrast to networks on a spatial structure. Here, we study dynamical and spectral properties of finite-sized chimeras on two-population networks. First, we elucidate how the Kuramoto order parameter of the finite sized globally coupled two-population network of phase oscillators is connected to that of the continuum limit. These findings suggest that it is suitable to classify the chimera states according to their order parameter dynamics, and therefore we define Poisson and Non-Poisson chimera states. We then perform a Lyapunov analysis of these two types of chimera states which yields insight into the full stability properties of the chimera trajectories as well as of collective modes. In particular, our analysis also confirms that Poisson chimeras are neutrally stable. We then introduce two types of ‘perturbation’ that act as small heterogeneities and render Poisson chimeras attracting: A topological variation via the simplest nonlocal intra-population coupling that keeps the network symmetries, and the allowance of amplitude variations in the globally coupled two-population network, i.e., we replace the phase oscillators by Stuart-Landau oscillators. The Lyapunov spectral properties of chimera states in the two modified networks are investigated, exploiting an ansatz based on the network symmetry-induced cluster pattern dynamics of the finite size network.
\end{abstract}
\maketitle

\begin{quotation}
Chimera states are a peculiar type of synchronization patterns in homogeneous oscillatory systems~\cite{pikovksy_sync,strogatz_sync} where regions of synchrony and asynchrony form spontaneously~\cite{kuramoto2002}. They were observed in diverse experiments~\cite{pendulum2,krischer1,exp1,exp2,exp3,exp4,exp5} and are believed to be important for certain biological manifestations, such as unihemispheric sleep of some animals or so-called bump-states of neural activity~\cite{neuro1,neuro2}. Also from a theoretical point of view, an understanding of chimera states plays an important role, as they mediate between order and disorder~\cite{omelchenko1,omelchenko2,abrams_review}. A detailed analysis of their dynamics is much facilitated with a simple topology, the simplest one consisting of two coupled populations~\cite{abrams_chimera2008,abrams_chimera2016,pazo_prx,Kurths_twogroup,basin,Laing_hetero,Laing_nonlocal,hetero-phaselag,timevarying_twogroup,Laing_SL2010,Laing_SL2019,pikovsky_WS}. For this minimal model, analytical results about the stability and bifurcations of chimera states could be obtained in the continuum limit~\cite{abrams_chimera2008}, and for the case of small populations it was shown the same type of bifurcations exist~\cite{abrams_chimera2016}. Yet, there are still many open questions, some of which we answer in this paper.

The incoherent dynamics of the two-population network depends sensitively on the initial conditions, and on the ensemble size~\cite{abrams_chimera2016,Kurths_twogroup,Laing_hetero}. In particular, when the initial conditions are obtained from the Poisson kernel, the incoherent motion is simpler than for general initial conditions~\cite{pikovsky-heteroWS,pikovsky_WS}. Our paper is centered around the questions how the chimera states can be classified according to the initial condition and how the dynamics of large and small size populations are linked. Another question we address is how to make the special chimera state with the simpler dynamics of the incoherent oscillators attracting in more realistic situations. Our analysis suggests the definition of a Poisson chimera which gives a natural way to classify the chimera states arising from different initial conditions. The main methods employed is Lyapunov analysis~\cite{CLV1,CLV2,CLV3,ipr,kevin,pikovsky_LE} and  network  symmetry~\cite{yscho,pecora1,pecora2}.

\end{quotation}

\section{\label{sec:introduction}Introduction}

Chimera states were first discovered for non-locally coupled phase oscillators on a spatially one-dimensional ring\cite{kuramoto2002}. To obtain a deeper understanding of the dynamics of chimera states, several mathematically more easily tractable models that still exhibit the primary dynamical properties of chimera states have been proposed~\cite{omelchenko1,omelchenko2,abrams_review}. The simplest of them is a network consisting of two populations of identical oscillators. All oscillators within one population are globally coupled to each other with a given intra-population coupling strength, which is the same for both populations. The coupling of the oscillators of different populations is all-to-all as well, but the inter-population coupling strength is different from the intra-coupling strength. In a chimera state of such a two-population topology one population oscillates fully synchronously while the other one exhibits incoherent oscillations. The network topology makes sure that the synchronized oscillators live on an invariant sync-manifold, which causes the simpler mathematical accessibility of these chimera states compared to those in other networks, e.g., on the spatially one-dimensional ring ~\cite{wolfrum_ring1,wolfrum_ring2}.

This simpler structure has been exploited in numerous studies~\cite{abrams_chimera2008,abrams_chimera2016,Laing_SL2010,Laing_SL2019,Laing_nonlocal,Laing_hetero,pazo_prx,pikovsky_WS,Kurths_twogroup,timevarying_twogroup,hetero-phaselag,pendulum1,pendulum2,basin}. In many of them the continuum limit was considered~\cite{abrams_chimera2008,pazo_prx,Kurths_twogroup}. Furthermore, in order to address the robustness of chimera states, heterogeneities have been introduced~\cite{Laing_hetero,hetero-phaselag,Laing_hetero2} or non-complete networks of oscillators were considered with a static~\cite{Laing_nonlocal} or time varying~\cite{timevarying_twogroup} network structure. Besides phase oscillators also planar oscillators were studied~\cite{Laing_SL2010,Laing_SL2019}.

Studies with finite sized populations revealed a strong dependence of the chimera states on initial conditions (ICs)~\cite{Kurths_twogroup,Laing_hetero,basin, pikovsky_WS,pikovsky-heteroWS,pikovsky-heteroWS2}. The simplest chimera dynamics was obtained when the ICs of the incoherent population were distributed according to the Poisson kernel. However, the chimera states in the identical phase oscillator model were shown to be neutrally stable in many directions~\cite{pikovsky_WS}. In contrast, when heterogeneous populations were considered, the asymptotic dynamics even for slightly off Poisson kernel ICs was found to be attracting in the long time limit~\cite{OA2,pikovsky-heteroWS, Laing_SL2010}.

In the following we will term such ICs Poisson initial conditions and abbreviate them with $\textbf{PIC}$, whereas all other initial conditions are referred to as non-Poisson ICs and abbreviated by $\textbf{n-PIC}$. In the case of $\textbf{PIC}$s, the chimera states of small-sized populations exhibited pronouncedly different order parameter dynamics from large-sized populations, which has been attributed to finite-size fluctuations~\cite{abrams_chimera2016}. Moreover, for large populations, the numerical simulation suggested that the order parameter becomes indistinguishable from the one predicted by the continuum limit~\cite{abrams_chimera2008,abrams_chimera2016,Kurths_twogroup}. 

\begin{figure}[t!]
\includegraphics[width=1.0\linewidth]{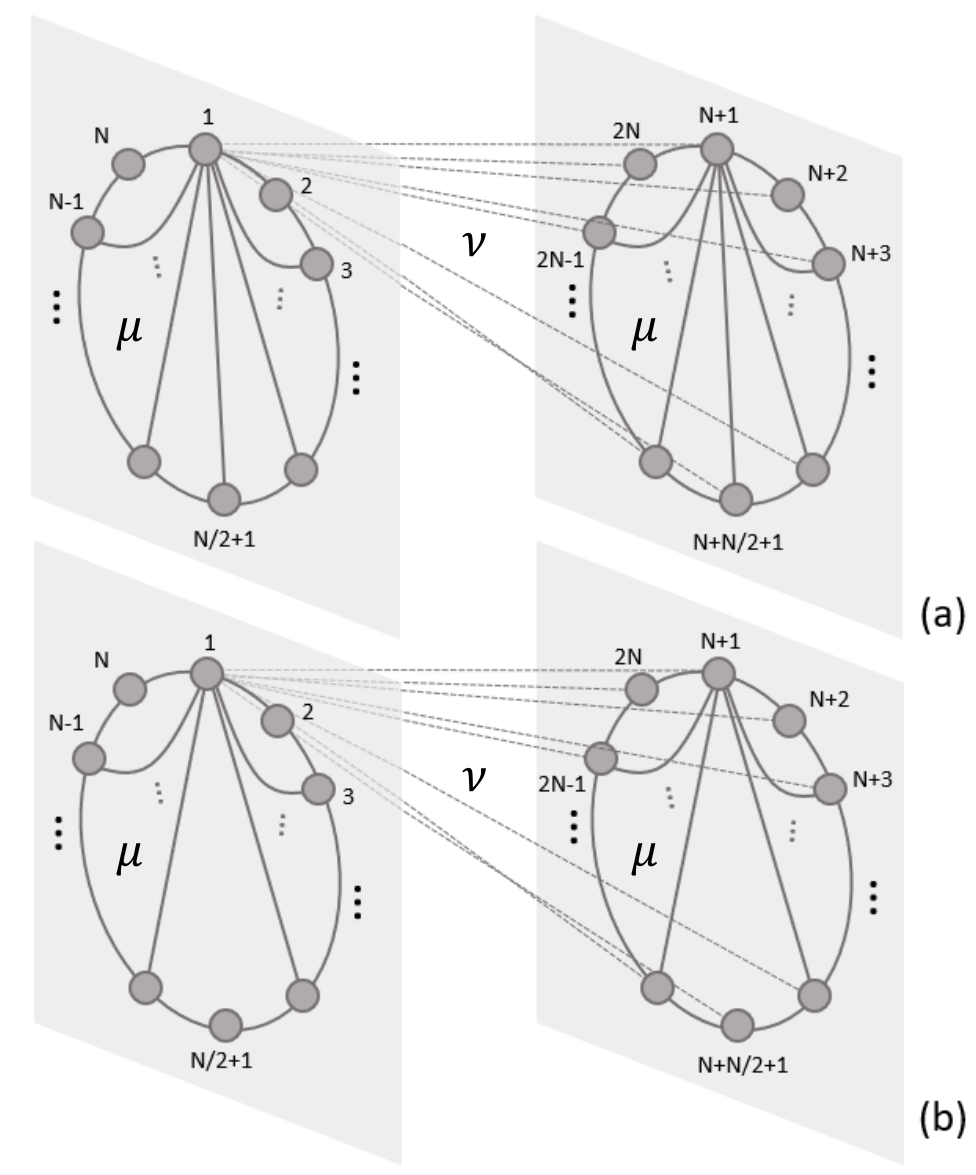}
\caption{Schematics of the two-population network topologies considered in this paper. (a) Global intra- and inter- population topology, and (b) global inter- and nonlocal intra-population coupling. Here, only the connections from the first oscillator are fully depicted. The solid connections indicate the intra-population coupling with  strength $\mu$, and the dashed one the inter-population connections with  strength $\nu$. Note that in the nonlocal intra-population topology, each oscillator is connected to all the other oscillators except of the opposite one. } 
\label{Fig:network}
\end{figure}

In this paper, we elucidate the origin of both the impact of the initial conditions and of the population size on the chimera dynamics in two-population networks. In particular, we present evidence that there is a continuous change from the small to the large size populations up to the continuum limit. First, we consider the classical two-population network topology with identical Kuramoto-Sakaguchi phase oscillators and global intra- and inter-population coupling (Fig.~\ref{Fig:network} (a)). We demonstrate that finite-sized chimeras emerging from $\textbf{PIC}$ live in the neutrally stable Poisson submanifold, which corresponds to the Ott-Antonsen (OA) manifold in the continuum limit and on which the incoherent phase degrees of freedom (DOFs) are distributed according to the Poisson kernel~\cite{mobius_strogatz,pikovsky_WS,Laing_SL2010}. To underline the different dynamical characteristics of chimera states arising from $\textbf{PIC}$s and $\textbf{n-PIC}$s,  we introduce the concept of a Poisson chimera trajectory and illustrate that what has been so far considered as finite-size fluctuations of small-size chimeras is of fundamentally different nature in the case of Poisson chimeras and of chimeras resulting from $\textbf{n-PIC}$s.   

As the next step, we introduce two simple ways that render such Poisson chimera states stable in the sense that they attract nearby trajectories that start from $\textbf{n-PIC}$ or at least evolve towards a close vicinity of the Poisson submanifold~\cite{pikovsky-heteroWS2,OA2,OA-attractive,pikovsky-heteroWS,pikovsky-heteroWS3}. The first approach introduces a small topological perturbation of the network structure which leads to the simplest nonlocal intra-population coupling that is represented by a specific adjacency matrix that preserves the network symmetry as the system size increases (Fig.~\ref{Fig:network} (b)). Then, we allow for amplitude degrees of freedom (DOFs) by coupling Stuart-Landau oscillators instead of phase oscillators~\cite{Laing_SL2010,Laing_SL2019}. Here, both the global and the nonlocal intra-population network topologies are used.
 
Our main method to access the properties of the various chimera trajectories is Lyapunov spectral analysis, which yields the spectra of the Lyapunov exponents (LEs) and the covariant Lyapunov vectors (CLVs)~\cite{CLV1,CLV2,CLV3,ipr,kevin}. The analysis reveals whether the incoherent oscillator population is attractive or not, as well as the full stability information of the synchronized population. In order to analytically address and approximate the Lyapunov exponents, an approach is introduced that is based on the network symmetry-induced cluster pattern analysis~\cite{yscho,pecora1,pecora2}. Here, we exploit the fact that the finite-sized two-population topology can be viewed as one network that possesses the inherent network symmetries represented by the automorphism group\cite{kudose,macarthur,yscho2}. The details of the background theories are compiled in Appendices ~\ref{append:lyapunov}-\ref{append:network-symmetry}.

The rest of this paper is organized as follows. In Sec.~\ref{sec:poisson-random-chimera}, we investigate the properties of chimera states of phase oscillators according to the initial conditions and define Poisson chimeras as opposed to non-Poisson chimeras. Furthermore, we discuss the Lyapunov spectral properties of these chimeras. In Sec.~\ref{sec:attracing-Poisson}, we consider two ways that render Poisson chimeras attractive; nonlocal topology and amplitude variables. Finally, we summarize the results in Sec.~\ref{sec:conclusion}.

\section{\label{sec:poisson-random-chimera} Poisson and non-Poisson Chimeras}

\subsection{\label{subsec:governing model} Model and Observable Dynamics}

In this section, we consider a set of identical Kuramoto-Sakaguchi (KS) phase oscillators arranged in the two-population network topology with global inter- and intra-population coupling of different strengths as depicted in Fig.~\ref{Fig:network} (a). This system is considered to be the simplest model that exhibits chimera states coexisting with a stable complete synchronization state.~\cite{abrams_chimera2008,abrams_chimera2016}.

\begin{figure}[t!]
\includegraphics[width=1.0\linewidth]{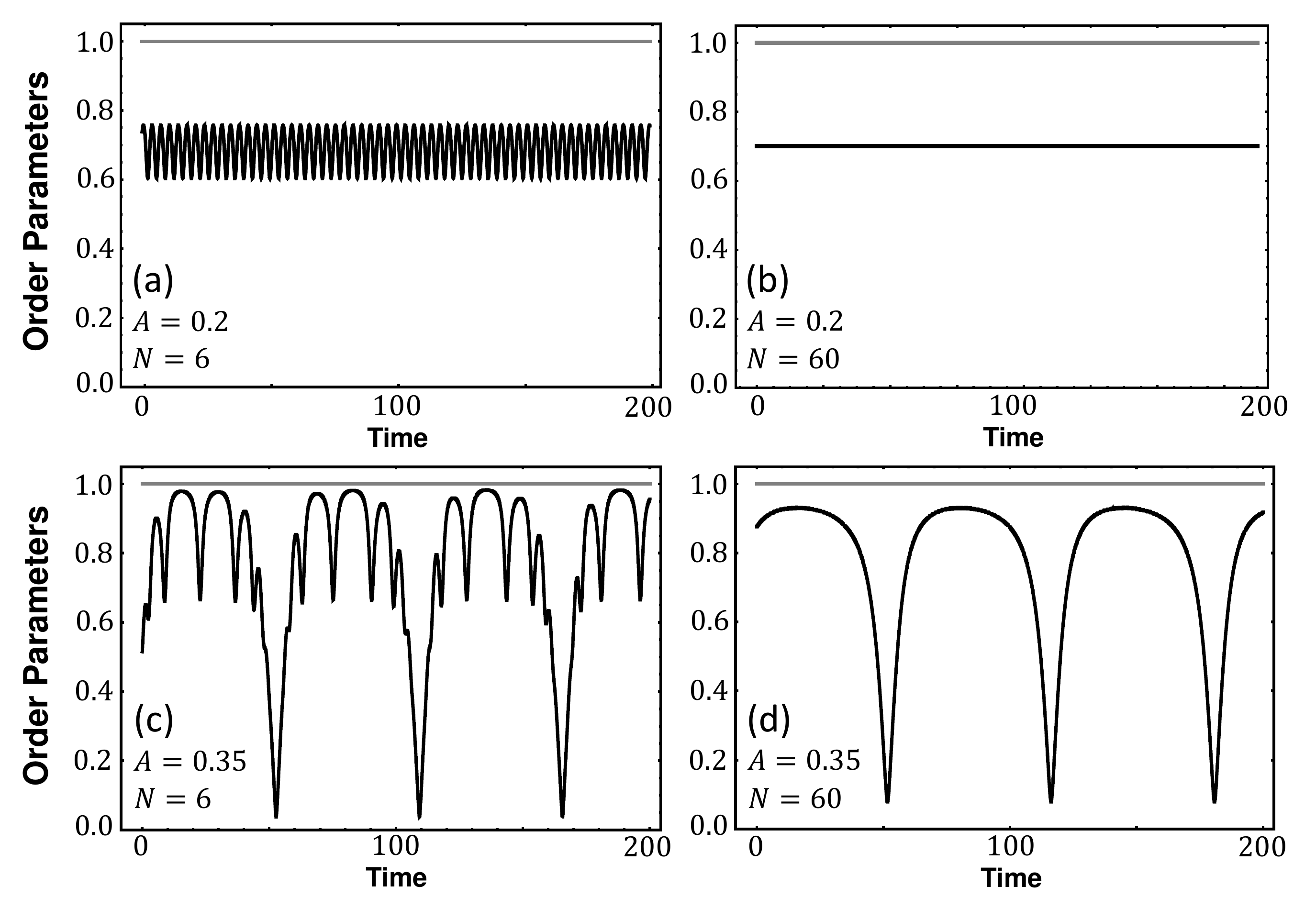}
\caption{The magnitudes of Kuramoto order parameters $r(t)$ of the coherent and incoherent populations of chimera states in the two-population network starting from \textbf{PICs} after transients have died out ($t \geq 10^5$). For each figure, the gray solid line indicates the order parameter for the perfectly synchronized population $(r(t)=1)$ and the black solid line the incoherent population $(r(t)<1)$: (a-b) Stationary chimera states with $A=0.2$ and (c-d) breathing chimera states with $A=0.35$ for the system sizes $N=6$ (left) and $N=60$ (right), respectively. } 
\label{Fig:KS-Poisson-chimera-orderparameter}
\end{figure}

Each of the two interacting populations is composed of $N$ phase oscillators. The state of each oscillator is fully described by its phase $\phi_i \in \mathbb{T} = [-\pi, \pi)$ for $i=1,...,2N$. The governing equations of the oscillators in the first population are
\begin{flalign}
\frac{d\phi_i(t)}{dt} &= \omega + \frac{\mu}{N}\sum_{j=1}^{N}\textrm{sin}(\phi_j(t)-\phi_i(t)-\alpha) \notag\\&+\frac{\nu}{N}\sum_{j=1}^{N}\textrm{sin}(\phi_{j+N}(t)-\phi_i(t)-\alpha)
\label{Eq:KS-global-governing1}
\end{flalign}
with $i=1,...,N$, and those of the second population are
\begin{flalign}
\frac{d\phi_{i+N}(t)}{dt} &= \omega + \frac{\mu}{N}\sum_{j=1}^{N}\textrm{sin}(\phi_{j+N}(t)-\phi_{i+N}(t)-\alpha)\notag \\ &+\frac{\nu}{N}\sum_{j=1}^{N}\textrm{sin}(\phi_{j}(t)-\phi_{i+N}(t)-\alpha)
\label{Eq:KS-global-governing2}
\end{flalign}
with $i=1,...,N$.  Notice that all the oscillators are identical, i.e., they have the same intrinsic frequency $\omega=0$ and the same Sakaguchi phase-lag parameter $\alpha=\pi/2-\beta$ where $\beta$ is small enough such that chimera states exist~\cite{abrams_review,omelchenko1}. $\nu$ and $\mu$ are the inter- and intra-population coupling strengths (see Fig.~\ref{Fig:network}). We rescale time such that $\mu+\nu=1$ and define $A=\mu-\nu$. Throughout this work, we set $\beta =0.08$ and $A$ either $0.2$ or $0.35$. This choice of parameters yields chimera states that are representative of so-called stationary and breathing chimeras, respectively, which are characterized by a stationary and oscillatory behavior of the magnitude of the Kuramoto order parameter with time for large populations~\cite{abrams_chimera2008}. The Kuramoto order parameters for the two populations are defined by $r_1(t) e^{i \Theta_1(t)} = \frac{1}{N}\sum_{j=1}^{N}e^{i \phi_j(t)}$ and $r_2(t) e^{i \Theta_2(t)} = \frac{1}{N}\sum_{j=1}^{N}e^{i \phi_{j+N}(t)}$. Chimera states in a two-population network have one population consisting of perfectly synchronized oscillators with $r_{\textrm{sync}}(t)=1$ and the other one being composed of incoherent oscillators with $0<r_{\textrm{incoh}}(t)<1$.~\cite{abrams_chimera2016}.

\begin{figure}[t!]
\includegraphics[width=1.0\linewidth]{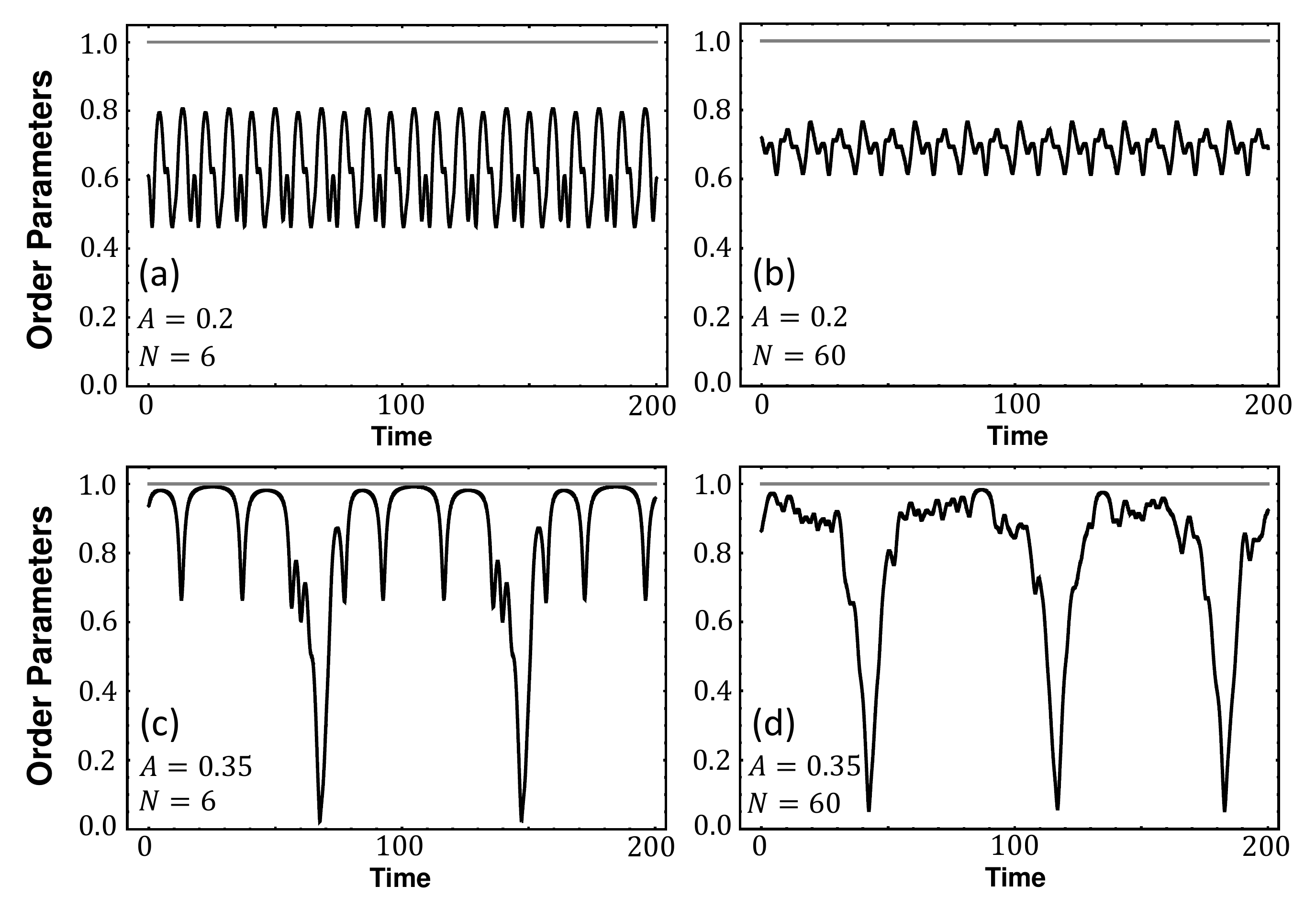}
\caption{The magnitudes of Kuramoto order parameters $r(t)$ of the coherent and incoherent populations of chimera states in the two-population network starting from $\textbf{n-PICs}$ after transients have died out ($t \geq 10^5$). For each figure, the gray solid line indicates the order parameter for the perfectly synchronized population $(r(t)=1)$ and the black solid line the incoherent population $(r(t)<1)$: (a-b) $A=0.2$ (for which with $\textbf{PICs}$ stationary chimeras are obtained) and (c-d) $A=0.35$ (for which with $\textbf{PICs}$ breathing chimeras are obtained) for the system sizes $N=6$ (left) and $N=60$ (right), respectively. } 
\label{Fig:KS-small-size Random chimera}
\end{figure}

Numerical solutions of  Eqs.~(\ref{Eq:KS-global-governing1}-\ref{Eq:KS-global-governing2}) suggest that for each parameter set $A$ and $\beta$ the chimera trajectories can be divided into two groups, depending on the initial conditions. If the trajectory starts from \textbf{PICs} (the detailed description of ICs will be given in the next section), a chimera trajectory shows a simple, regular motion of the magnitude of the order parameter as depicted in Fig.~\ref{Fig:KS-Poisson-chimera-orderparameter}. For large population numbers $N$ as in Fig.~\ref{Fig:KS-Poisson-chimera-orderparameter} (b,d), the magnitude of the order parameter $r_{\textrm{incoh}}(t)$ of chimera states emerging from \textbf{PICs} is either stationary in time (Fig.~\ref{Fig:KS-Poisson-chimera-orderparameter} (b)) or exhibits simple periodic oscillations (Fig.~\ref{Fig:KS-Poisson-chimera-orderparameter} (d)), depending on the value of $A$. These dynamics were termed stationary and breathing chimeras, respectively~\cite{abrams_chimera2008}, and $r_{\textrm{incoh}}(t)$ is virtually indistinguishable from the one of the OA solution in the continuum limit. For small population sizes $N$, as in Fig.~\ref{Fig:KS-Poisson-chimera-orderparameter} (a,c), $r_{\textrm{incoh}}(t)$ is composed of two contributions: the motion it shows in the case of large $N$ and a superposed, in the case of breathing chimeras secondary, oscillation. Note that throughout this paper, we name each chimera state according to its classification in the continuum limit at the given parameter set for the sake of simplicity. When the chimera trajectory starts from \textbf{n-PICs}, in contrast, $r_{\textrm{incoh}}(t)$ shows a more complicated motion, strongly depending on the given initial conditions (Fig.~\ref{Fig:KS-small-size Random chimera}). This initial condition dependence of $r_{\textrm{incoh}}(t)$ has been pointed out previously~\cite{Kurths_twogroup,pikovsky_WS,pikovsky-heteroWS2,abrams_chimera2008}, and it has led many authors to use rather special initial conditions for their chimera studies. In this work, we will address the initial condition dependence in some detail, and introduce the concept of Poisson and non-Poisson chimeras in the next section. Furthermore, we explaining the stability of both synchronized and incoherent populations with a Lyapunov analysis.

\subsection{\label{subsec:poisson-ch}\label{subsec:random-ch}Poisson and Non-Poisson Chimeras}

As mentioned above, in order to obtain the simple motion of the magnitude of the order parameter as depicted in Fig.~\ref{Fig:KS-Poisson-chimera-orderparameter} and also in Refs.~\onlinecite{abrams_chimera2008,abrams_chimera2016}, a specific initial condition has to be used. We coin this initial condition \textit{Poisson initial condition} ($\textbf{$\textbf{PIC}$}$) since the initial incoherent phases are generated from the Poisson kernel that corresponds to the OA manifold in the continuum limit~\cite{mobius_strogatz,OA,OA2,Laing_hetero}. To obtain $\textbf{PICs}$, one first has to solve the 2-dimensional Ott-Antonsen reduced equations for the incoherent population, which for the stable stationary chimera state with the parameter set $A=0.2$ and $\beta=0.08$ results in $\rho_0=0.69998$ and $\varphi_0=6.11918$, where $\varphi_0 = \varphi_1 - \varphi_2$ and $\varphi_i$ for $i=1,2$ is the OA phase variable for each population, respectively~\cite{abrams_chimera2008}. Then, consider the Poisson kernel
\begin{flalign}
f^{(2)}(\phi;\rho_0,\varphi_0) &=  \frac{1}{2\pi} \Bigg[ 1+  \sum_{n=1}^{\infty} \bigg( \big( a_0 e^{i\phi} \big)^n + \textrm{c.c.} \bigg) \Bigg] \notag \\
    &= \frac{1}{2\pi} \frac{1-\rho_0^2}{1-2 \rho_0 \textrm{cos}(\phi-\varphi_0)+\rho_0^2}
    \label{Eq:poisson-kernel}
\end{flalign} where $a_0=\rho_0 e^{-i\varphi_0}$, and its inverse cumulative distribution function (inverse CDF). For our finite-size chimeras, we want the initial incoherent phase distribution $\{\phi_{i+N}(0)\}_{i=1}^{N}$ to be as close as possible to Eq.~(\ref{Eq:poisson-kernel}). To obtain such ICs, equally spaced probabilities are used as arguments of the inverse CDF of the Poisson kernel, i.e., $N$ initial phases of the incoherent population are numerically obtained from
\begin{flalign}
\frac{i-\frac{1}{2}}{N} &= \int_{-\pi}^{\phi_{i+N}(0)}  \frac{1}{2\pi} \frac{1-\rho_0^2}{1-2 \rho_0 \textrm{cos}(\phi-\varphi_0)+\rho_0^2} d\phi
\label{Eq:OA-I-C}
\end{flalign}
for $i=1,...,N$. For the synchronized population, the initial phases $\{ \phi_i(0) \}_{i=1}^N$ are picked from the delta distribution $f^{(1)}(\phi)=\delta(\phi-\phi_0)$ which manifests that this population consists of the perfectly synchronized oscillators.

Simulations of the governing  Eqs.~(\ref{Eq:KS-global-governing1}-\ref{Eq:KS-global-governing2}) can also be initiated from an $\textbf{n-PIC}$. In this work, \textbf{n-PIC} consists of initial phases $\{ \phi_{i}(0) \}_{i=1}^{2N}$ that are randomly and independently from each other picked from the uniform distribution within $[-\pi,\pi)$ . Note that such initial conditions do not cover the entire manifold of the incoherent oscillator population off the Poisson submanifold but rather only correspond to some subset of the entire manifold corresponding to the incoherent population.

As we have pointed out above, starting from $\textbf{PIC}$s, the magnitude of the order parameter exhibits one of two behaviors, depending on the population size $N$.  For large $N$, $r_{\textrm{incoh}}(t)$ is virtually indistinguishable from the one of the continuum limit which is a solution of the OA reduced dynamics~\cite{abrams_chimera2008}. For small $N$, the motion of $r_{\textrm{incoh}}(t)$ is comprised of the main motion close to the OA dynamics superimposed by a regular secondary oscillation. Its clear and regular behavior suggests that the small-size behavior is not just a finite-size fluctuation~\cite{abrams_chimera2016} but rather has a deterministic origin. In the following, we disclose the source of the secondary motion of $r_{\textrm{incoh}}(t)$ of small-size chimeras that start from $\textbf{PIC}$.

\begin{figure}[t!]
\includegraphics[width=1.0\linewidth]{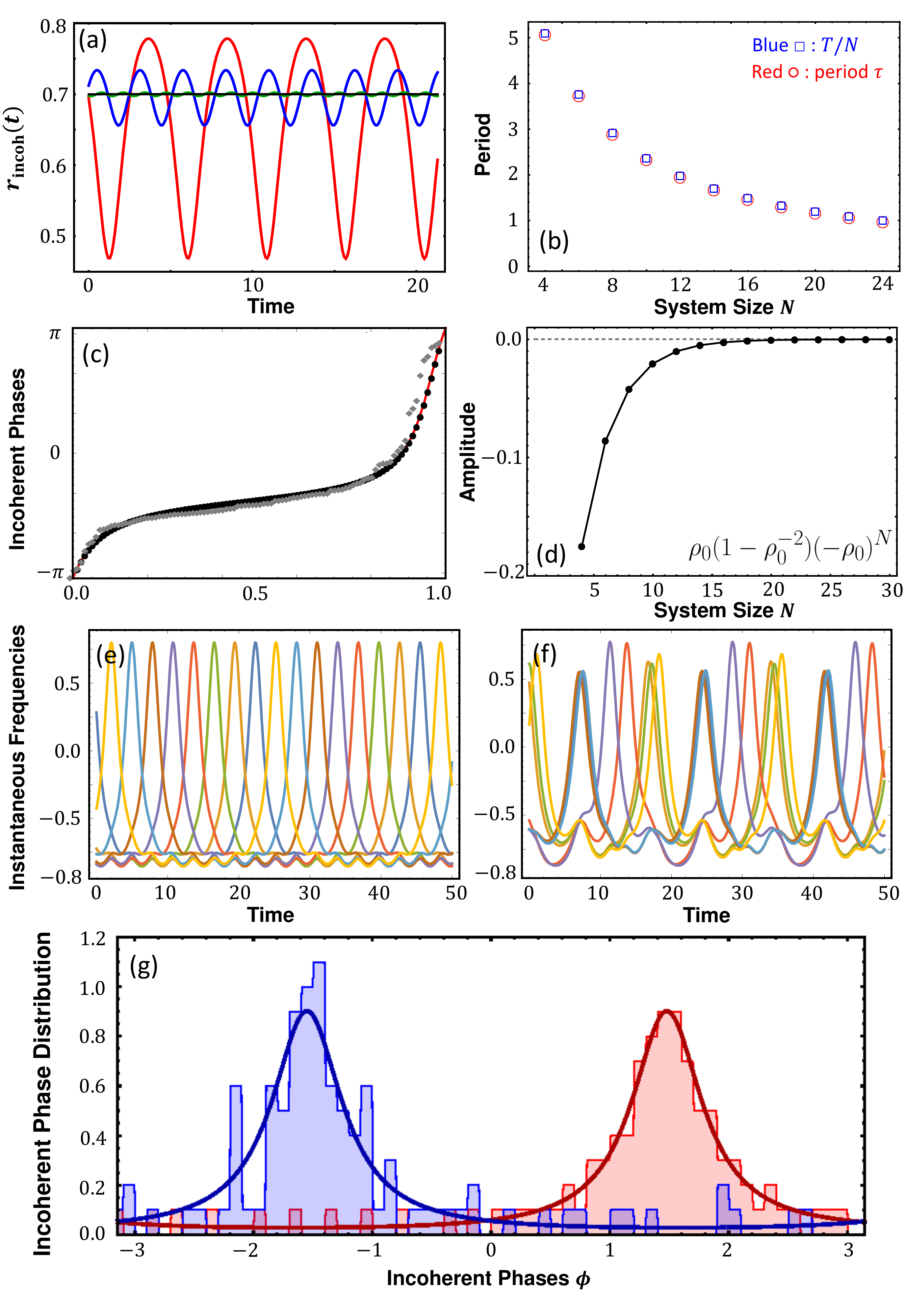}
\caption{(a) Oscillations of the magnitude of the order parameter for $A=0.2$ and different system sizes: $N=4$ (red), $N=8$ (blue), $N=16$ (green), and $N=32$ (black). (b) Period of $r_{\textrm{incoh}}(t)$ as determined numerically (red) and predicted from Eq.~(\ref{Eq:periodic-incoh-order}) as a function of the system size $N$. 
(c) Snapshot of the sorted incoherent phases in the numerical order with $N=100$ as a function of the rescaled index after a time $t \geq 10^6$ for a Poisson chimera (black dots), a non-Poisson chimera (gray diamonds) and the theoretical curve of the inverse CDF of Poisson kernel (red solid curve).  (d) Magnitude of the secondary oscillation as a function of  system size. (e,f) Instantaneous frequencies of the incoherent oscillators of the system $N=8$ for a Poisson chimera (e) and a non-Poisson chimera (f). (g) Snapshot of the incoherent phase distribution for a Poisson chimera (red) and the non-Poisson chimera (blue) for $t \geq 10^6$ with $N=100$ oscillators. Each solid line indicates the theoretical Poisson kernel curve corresponding to $\rho_0$ within an appropriate rotating frame. } 
\label{Fig:KS-small-size Poisson chimera}
\end{figure}

To address the dynamical behavior of the small-size chimeras, we first focus on the stationary chimera states with $A=0.2$.
Numerical integration of Eqs.~(\ref{Eq:KS-global-governing1}-\ref{Eq:KS-global-governing2}) with $\textbf{PIC}$s reveals that the instantaneous velocity of each incoherent oscillator $\{\dot{\phi}_{i+N}(t)\}_{i=1}^{N}$ is in fact a periodic function, and, furthermore, all instantaneous velocities of the incoherent oscillators have the same functional form and share the same period $T$. On the level of the instantaneous velocities this behavior is reminiscent of the behavior of the instantaneous phases in a splay state~\cite{sjlee1,splay1,splay2} (see Fig.~\ref{Fig:KS-small-size Poisson chimera} (e)). Numerically, the period of the instantaneous velocity $T$ has a value $T \approx 23.48$, irrespective of the population size $N$. Hence, we assume that the instantaneous frequencies of the incoherent oscillators have the form of a splay state such that $\dot{\phi}_{i}(t-\frac{j}{N}T)=\dot{\phi}_{i+j}(t)$ for an arbitrary $j \in \{1,...,N\}$, which gives $\phi_{i}(t-\frac{1}{N}T) = \phi_{i+1}(t)+W$ for $i=N+1,...,2N$ with $\phi_{2N+1} \equiv \phi_{N+1}$ where $W \in \mathbb{R}$ is a common constant. Plugging the expression for $\{\phi_{i+N}(t)\}_{i=1}^{N}$ in the definition of the order parameter, we obtain 
\begin{flalign}
r_{\textrm{incoh}}(t) &=  \bigg|\frac{1}{N} \sum_{k=N+1}^{2N} e^{i \phi_{k+1}(t)} \bigg| =  \bigg|\frac{1}{N} \sum_{k=N+1}^{2N} e^{i(\phi_{k}(t-\frac{T}{N})-W)} \bigg|   \notag \\ 
 &= \bigg|\frac{e^{-i W}}{N} \sum_{k=N+1}^{2N} e^{i \phi_{k}(t-\frac{T}{N})} \bigg|=  \bigg|\frac{1}{N} \sum_{k=N+1}^{2N} e^{i \phi_{k}(t-\frac{T}{N})} \bigg| \notag \\ &= r_{\textrm{incoh}}\bigg(t-\frac{T}{N}\bigg) = r_{\textrm{incoh}}(t-\tau)
\label{Eq:periodic-incoh-order}
\end{flalign}
for all $t \in \mathbb{R}$. Thus, $r_{\textrm{incoh}}(t)$ in Eq.~(\ref{Eq:periodic-incoh-order}) is indeed a periodic function and its period $\tau=T/N$ is continuously decreasing as $N$ increases. In Fig.~\ref{Fig:KS-small-size Poisson chimera} (b), the numerical calculations of the period of the order parameter are plotted as a function of $N$ together with the values predicted by Eq.~(\ref{Eq:periodic-incoh-order}). The nearly perfect agreement of both values confirms that the period $\tau(N)$ of the order parameter oscillations are indeed decreasing with $N$ according to $T/N$. 

Next, we investigate the amplitude of the periodic order parameter of a small-size stationary chimera. As obvious from Fig.~\ref{Fig:KS-small-size Poisson chimera} (a), the amplitude of $r_{\textrm{incoh}}(t)$ also decreases with increasing $N$. To explain this, we here consider the Watanabe-Strogatz reduced dynamics $\rho_2(t)$, $\Phi_2(t)$, and $\Psi_2(t)$ for the incoherent population~\cite{WStheory,abrams_chimera2016}. These quantities are related to the Kuramoto order parameter according to\cite{pikovsky_WS,pikovsky-heteroWS,abrams_chimera2016}
\begin{equation}
    r_{\textrm{incoh}}(t)e^{i \Theta_{\textrm{incoh}}(t)} = \rho_2(t) e^{i \Phi_2(t)}\gamma_2(\rho_2,\Psi_2)
    \notag
\end{equation}
where
\begin{equation}
   \gamma_2(\rho_2,\Psi_2) = \frac{1}{N \rho_2} \sum_{k=1}^{N}\frac{\rho_2+e^{i(\psi^{(2)}_k-\Psi_2)}}{1+\rho_2 e^{i(\psi^{(2)}_k-\Psi_2)}} \label{Eq:gamma-definition}
\end{equation} and $\{\psi_k^{(2)} \}_{k=1}^N$ are the constants of motion, which are determined by the given initial conditions and satisfy three appropriate constraints~\cite{pikovsky-heteroWS}. For $\textbf{PIC}$s, the constants of motion comply with the uniform distribution $\psi_k^{(2)} = \frac{2\pi k}{N}$ for $k=1,...,N$~\cite{abrams_chimera2016,pikovsky_WS}. For $\gamma_{2}$ one can obtain~\cite{pikovsky-heteroWS,pikovsky-heteroWS2}
\begin{equation}
\gamma_{2} = 1 + (1-\rho_2^{-2})(-\rho_2)^N  \frac{ e^{i N(\frac{2\pi}{N}-\Psi_2)}}{1-(-\rho_2)^N e^{i N(\frac{2\pi}{N}-\Psi_2)} }.
\label{Eq:WS-gamma}
\end{equation}
Numerical calculations suggest that the values of the radial variable $\rho_2(t)$ are consistent with the stationary OA radial variable, 
while exhibiting a very small finite-size oscillation that in this context we can ignore, even for the smallest chimera.
Hence, we can assume $\rho_2(t)=\rho_0$ in Eq.~(\ref{Eq:WS-gamma}). Then, the Kuramoto order parameter can be rewritten as
\begin{flalign}
r_{\textrm{incoh}}(t) &= \rho_0 |\gamma_2(t)| = \rho_0 \Bigg| \frac{1}{N \rho_0} \sum_{k=1}^{N}\frac{\rho_0+e^{i(\frac{2\pi k}{N}-\Psi_2)}}{1+\rho_0 e^{i(\frac{2\pi k}{N}-\Psi_2)}} \Bigg| \notag \\ &= \bigg| \rho_0 + \rho_0 (1-\rho_0^{-2})(-\rho_0)^N \mathbf{O}(t;\rho_0,\Psi_2)  \bigg|
\label{Eq:order-gamma}
\end{flalign} where $\mathbf{O}(t;\rho_0,\Psi_2) = \frac{e^{i N(\frac{2\pi}{N}-\Psi_2(t))} }{1-(-\rho_0)^N e^{i N(\frac{2\pi}{N}-\Psi_2(t))} }$. The second term in Eq.~(\ref{Eq:order-gamma}) represents the secondary oscillation of the small-size stationary chimeras. In Fig.~\ref{Fig:KS-small-size Poisson chimera} (d), the amplitude of the secondary oscillation is plotted as a function of $N$. It decreases monotonically with $N$ and approaches zero as $N \rightarrow \infty$. Thus, the periodic behavior of $r_{\textrm{incoh}}(t)$ gradually disappears with increasing $N$, such that $r_{\textrm{incoh}}(t) \rightarrow \rho_0$ as $N \rightarrow \infty$. Regarding the small-size breathing chimera state, $r_{\textrm{incoh}}(t)$ shows the main breathing motion while having the small secondary oscillation along it. It depends on the system size in a similar manner as the stationary chimeras do, namely according to
\begin{equation}
    r_{\textrm{incoh}}(t) =\rho_2(t) \Bigg| 1+  (1-\rho^{-2}_{2}(t))(-\rho_2(t))^N \mathbf{O}(t;\rho_2(t),\Psi_2(t)) \Bigg| \notag
\end{equation}
where $\rho_2(t)$ is no longer a fixed constant but exhibits the main breathing motion (see Fig.~\ref{Fig:KS-Poisson-chimera-orderparameter} (c)). As in the case of the stationary chimeras, the secondary oscillation vanishes for sufficiently large system sizes since $\rho(t)<1$ for $\forall t \geq 0$, which makes $(1-\rho_2(t))(-\rho_2(t))^N \rightarrow 0$ as $N \rightarrow \infty$ and the dynamics of the chimera states approach the one of the continuum limit.

Our analysis has revealed that both period and amplitude of the secondary oscillation of $r_{\textrm{incoh}}(t)$ continuously decrease as the system size increases. From approx. $N \gtrsim 24$ on, the secondary oscillation is not discernible anymore. Rather, $r_{\textrm{incoh}}(t)$  displays a motion indistinguishable from the one of the OA dynamics in the continuum limit. We therefore classify chimeras with population sizes  $N \gtrsim 24$ as large-size chimeras, those with $N < 24$ as small size chimeras. Yet, we would like to point out that there is a continuous change from the small-size to the large size chimeras and eventually up to the OA dynamics in the continuum limit as $N \rightarrow \infty$.  

On the other hand, when the chimeras started from \textbf{n-PIC}, a non-Poisson initial condition determines nonuniform constants of motion in the WS reduced dynamics. Then the stationary chimera states obtained from a given $\textbf{n-PIC}$ with the same parameter set ($A=0.2$ and $\beta$=0.08) show incoherent motion that is qualitatively different from the Poisson chimeras and depend on the specific initial conditions used, i.e., on the nonuniform constants of motion. Fig.~\ref{Fig:KS-small-size Random chimera} shows the temporal evolution of the magnitude of the order parameter for $\textbf{n-PIC}$s and otherwise identical parameter values and system sizes as Fig.~\ref{Fig:KS-Poisson-chimera-orderparameter} does for $\textbf{PIC}$s. Clearly, the behavior of $r_{\textrm{incoh}}(t)$ is more complicated in all four cases. In particular, the fluctuations of $r_{\textrm{incoh}}(t)$ do not disappear for the large-size chimeras and the overall motion of $r_{\textrm{incoh}}(t)$ of small-size chimeras is not composed of a superposition of the OA dynamics and the secondary oscillation. This is in line with the observation that the instantaneous velocities of the incoherent oscillators $\{ \dot{\phi}_{i+N}(t) \}_{i=1}^{N}$ do not form a splay state-like behavior 
but rather their shapes differ from oscillator to oscillator and the maxima are time-shifted by different amounts (Fig.~\ref{Fig:KS-small-size Poisson chimera} (f)). 
Notice that the quasiperiodic chimera states observed in Refs.~\onlinecite{pikovsky_WS,pikovsky-heteroWS,pikovsky-heteroWS2} are specific examples of non-Poisson chimera trajectories using a specific non-Poisson initial condition, or corresponding nonuniform constants of motion.

Finally, the red distribution in Fig.~\ref{Fig:KS-small-size Poisson chimera} (g)  illustrates that if the chimera trajectory starts from $\textbf{PIC}$, then the incoherent phase distribution of this chimera state remains in the Poisson kernel as defined in Eq.~(\ref{Eq:poisson-kernel}) within an appropriate rotating reference frame. This is confirmed by the observation that the incoherent phases sorted by their magnitude and plotted against its index (normalized to the total number of oscillators) coincide with the inverse CDF of Eq.~(\ref{Eq:poisson-kernel}) (Fig.~\ref{Fig:KS-small-size Poisson chimera} (c), black dots). This observation is consistent with the fact that the OA manifold is invariant under the dynamics in the continuum limit~\cite{mobius_strogatz,OA,OA2}. For the finite-sized chimeras initially starting from \textbf{PIC}, we can deduce from the splay form of $\dot{\phi}_{i}(t-\tau)=\dot{\phi}_{i+1}(t)$ that at least at $t=n\tau$ for $n\in \mathbb{N}$, the phases of the incoherent population are distributed according to the inverse CDF of the Poisson kernel since the splayed phase velocities result in the same constant shift for all the incoherent phases $\phi_{i}(t-\tau) = \phi_{i+1}(t)+W$. Beyond that, the numerical results indicate that the finite-sized Poisson submanifold along the chimera state starting from \textbf{PIC} is invariant under the dynamics. For example, let us define $E(t)=\Big| \langle e^{i \phi(t)} \rangle^2 - \langle e^{2 i\phi(t)} \rangle  \Big|$ where $\langle \cdot \rangle$ is the ensemble average, then for large enough $N$, $E(t)$ of the chimera trajectory starting from \textbf{PIC} is numerically found to be close to zero (more precisely, $E(t) \sim \mathcal{O}(10^{-5})$) revealing that the incoherent phases of such chimeras remain in the Poisson kernel. However, the large-size chimeras initiated from \textbf{n-PICs} do not have the incoherent phase distribution that satisfies the Poisson kernel (see Fig.~\ref{Fig:KS-small-size Poisson chimera} (c,g)), and after a long enough transient time $E(t) \sim \mathcal{O}(10^{-1})$. Thus, such chimera states initiated from $\textbf{n-PIC}$ should definitely be distinguished from the Poisson chimeras. Notice that the incoherent motion of the breathing chimera with $A=0.35$ starting from $\textbf{n-PIC}$ is different from the incoherent motion of the Poisson chimeras, and also depends on the given $\textbf{n-PIC}$ (see Fig.~\ref{Fig:KS-small-size Random chimera} (c-d)).

According to the above results we define a \textit{Poisson chimera trajectory} in the two-population network topology as follows: A chimera trajectory is a Poisson chimera if the phase DOFs $\{ \phi_{i}(t) \}_{i=1}^{2N}$ of a given ensemble of oscillators satisfy the following three dynamical characteristics: 
\begin{mydef}
The sync-population is perfectly synchronized and invariant.
\end{mydef}
\begin{mydef}
The incoherent phase distribution of Poisson chimeras remains in the Poisson kernel or at least in a close vicinity of the Poisson submanifold.
\end{mydef}
\begin{mydef}
Large-size Poisson chimeras are characterized by an incoherent order parameter being close to the one of the continuum limit, and the small-size Poisson chimeras by an incoherent order parameter whose motion is a superposition of the one of large-size Poisson chimeras and a secondary oscillation that continuously disappears through an increasing frequency and vanishing amplitude as $N \rightarrow \infty$.
\end{mydef}

Chimera states in the two-population network topology that do not fulfill \textbf{Conditions 1 - 3} are termed \textit{a non-Poisson chimera trajectory}. Note that the stationary Poisson chimera, whether small or large, has the additional property that the instantaneous frequencies of the incoherent oscillators are splayed within its period $T$ such that $\dot{\phi}_{i}(t-\frac{j}{N}T)=\dot{\phi}_{i+j}(t)$ for an arbitrary $j \in \{1,...,N\}$ and for $i=N+1,...,2N$ with $\phi_{2N+1}(t) \equiv \phi_{N+1}(t)$; however, the breathing chimeras do not.

For each parameter set, one can consider the manifold of the incoherent oscillator population. A state in this manifold can be characterized by $(N-3)$-parameter family of invariant subspaces determined by $N-3$ constants of motion, based on the WS framework (see Fig. 8 in Ref.~\onlinecite{WStheory}). The incoherent oscillators of the Poisson chimeras remain in the Poisson kernel, which corresponds to the Poisson submanifold (OA manifold in the continuum limit) in the following denoted by $\mathrm{M}_{\textrm{Poisson}}$ and the uniformly distributed constants of motion. However, the non-Poisson chimeras do not have such a property, corresponding to the invariant manifold outside of the Poisson submanifold denoted by $\mathrm{M}_{\textrm{incoh}}$ and general non-uniform constants of motion. In Ref.~\onlinecite{WStheory}, due to the constants of motion, the state for the identical oscillators described by the WS theory is neutrally stable in many directions. In the following, we will show the Lyapunov spectra in order to confirm the neutral stability of chimera states and then give two perturbations that render such chimera states attracting in the following sections.

\subsection{\label{subsec:LE-Poisson-and-nonPoisson-KS-global} Lyapunov Stability of Poisson and Non-Poisson Chimeras}

In this subsection, we investigate the stability of Poisson and non-Poisson chimeras. Therefore, we consider each chimera state as a reference trajectory in phase space and first numerically determine the Lyapunov exponents and then the corresponding covariant Lyapunov vectors. The properties of the resulting Lyapunov spectra are then elucidated using an ansatz based on network symmetry-induced cluster patterns~\cite{yscho,pecora1}. In particular, this method allows us to obtain approximate analytical expressions for the Lyapunov exponents associated with the synchronized population. Further insight into the Lyapunov exponents associated with the incoherent population is obtained from a Watanabe-Strogatz reduction of the dynamics. Finally, we present evidence of the existence of two collective modes. The detailed calculation for the synchronized population based on the network symmetry-induced cluster pattern dynamics is compiled in Appendix.~\ref{sec:mathematical-LE}.

\begin{figure}[t!]
\includegraphics[width=1.0\linewidth]{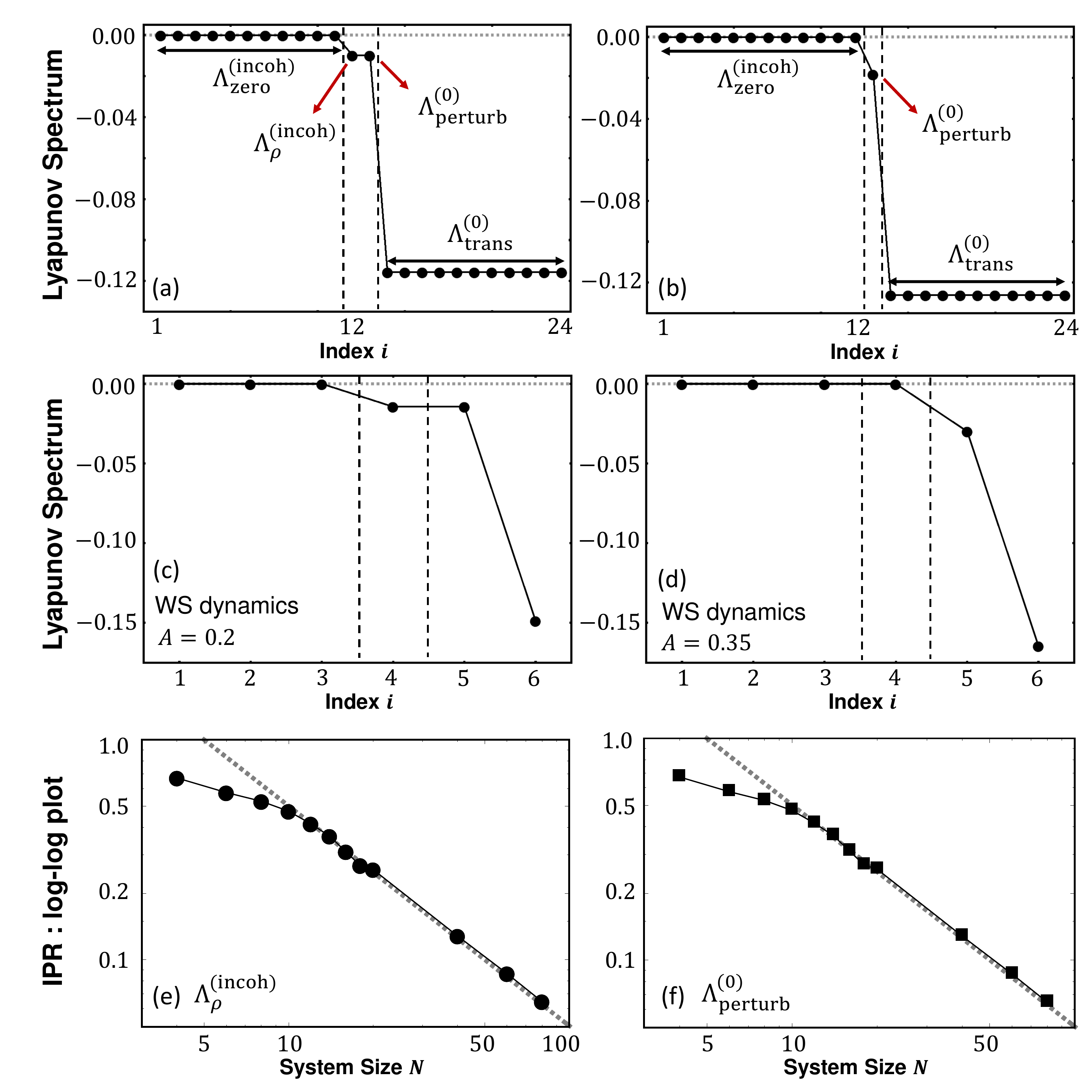}
\caption{(a-b) Lyapunov spectra of the full dynamics of Poisson chimera states with $N=12$ for $A=0.2$ (stationary chimeras) (a) and $A=0.35$ (breathing chimeras) (b). For the meaning of the $\Lambda$s see text. (c-d) Lyapunov spectra for the 6-dimensional Watanabe-Strogatz reduced dynamics of the chimera states in (a) and (b), respectively. The exponents marked by the black dashed lines indicate the LE corresponding to the radial WS variable. (e-f) Inverse Participation Ratio of the covariant Lyapunov vectors of the stationary chimera in (a) corresponding to $\Lambda_{\rho}^{(\textrm{incoh})}$ and $\Lambda_{\textrm{perturb}}^{(0)}$, respectively, as a function of system size $N$. The gray dashed lines indicates $\sim \frac{1}{N}$.} 
\label{Fig:KS-LE-Poisson}
\end{figure}

In Fig.~\ref{Fig:KS-LE-Poisson}, plates (a-b) display numerically determined Lyapunov spectra along Poisson chimera trajectories for stationary (a) and breathing (b) chimeras. Details about the numerical method used can be found in Refs.~\onlinecite{CLV1,CLV2,CLV3,kevin,ipr} and are summarized in Appendix \ref{append:lyapunov}. The Lyapunov spectrum of stationary chimera states (Fig.~\ref{Fig:KS-LE-Poisson} (a)) is composed of four groups of exponents: (i) $(N-1)$-fold degenerate zero exponents denoted by $\Lambda_{\textrm{zero}}^{(\textrm{incoh})}=0$, (ii) $(N-1)$-fold degenerate negative exponents denoted by $\Lambda_{\textrm{trans}}^{(0)}$, (iii) and (iv) two individual negative LEs, denoted by $\Lambda_{\textrm{perturb}}^{(0)}$ and $\Lambda_{\rho}^{(\textrm{incoh})}$. The spectrum obtained from a breathing Poisson chimera trajectory (Fig.~\ref{Fig:KS-LE-Poisson} (b)) exhibits a similar partition of the exponents, however, there is just one individual non-degenerate negative exponent, $\Lambda_{\textrm{perturb}}^{(0)}$, and the number of zero exponents has increased by 1 to $N$. These two type of partitions were characteristic for stationary and breathing Poisson chimeras, respectively, and independent of the system size $N$.

\subsubsection{\label{subsubsec:sync-LE-KS-global} Synchronized Population: $\Lambda_{\textrm{trans}}^{(0)}$ and $\Lambda_{\textrm{perturb}}^{(0)}$}

In Fig.~\ref{Fig:KS-LE-Poisson}, there are $(N-1)$-fold degenerate transverse Lyapunov exponents denoted by $\Lambda_{\textrm{trans}}^{(0)}$. The approximate analytical expressions of them are given as
\begin{equation}
\Lambda_{\textrm{trans},\kappa}^{(0)} =  -\mu \textrm{cos}\alpha -\frac{\nu}{N} \mathrm{Z}  < 0
\label{Eq:trans-LE-phase-global}
\end{equation} for $\kappa=2,...,N$ (indicating the indices for the $N-1$ transverse directions) where $\mathrm{Z} = \sum_{m'=1}^{N}\textrm{cos}(s_{m'}-s_0-\alpha)$ is treated as an external forcing field, and $\{s_m\}_{m=0}^{N}$ are the (coarse-grained) quotient dynamics of the chimera states according to the network cluster patterns discussed in Appendix.~\ref{sec:mathematical-LE}. The transverse Lyapunov exponents in Eq.~(\ref{Eq:trans-LE-phase-global}) are all negative and all degenerate, which confirms that the chimera state is stable in all directions transverse to the sync-manifold. Notice that the numerics ensures that $ -\mu \textrm{cos}\alpha \ll -\frac{\nu}{N} \mathrm{Z} < 0$. It also follows from numerical calculations that the covariant Lyapunov vectors corresponding to the LEs in Eq.~(\ref{Eq:trans-LE-phase-global}) have the form 
\begin{equation}
    \bold{v}_{\kappa}^{(0)} = [ v_{\kappa 1}^{(\textrm{trans})},...,v_{\kappa N}^{(\textrm{trans})},0,...,0]^\top \in \mathbf{T}_{\bm{\phi}_{\textrm{ch}}(t)}(\mathbb{T}^{2N}) \label{Eq:trans-CLV-KS-global}
\end{equation} for $\kappa=2,...,N$ where $\bm{\phi}_{\textrm{ch}}(t) \in \mathbb{T}^{2N}$ stands for the given chimera trajectory and $\mathbf{T}_{\bm{\phi}_{\textrm{ch}}(t)}(\mathbb{T}^{2N})$ is the tangent space at the point along such a chimera trajectory. These numerical CLVs have $\sum_{i=1}^{N}v_{\kappa i}^{(\textrm{trans})}=0$ which ascertains that these LEs correspond indeed to LEs transverse to the sync-manifold of the synchronized population.

We also discover in Fig.~\ref{Fig:KS-LE-Poisson} (a-b) another negative exponent $\Lambda_{\textrm{perturb}}^{(0)}$ for the synchronized population. The approximated value of it is given as
\begin{equation}
    \Lambda_{\textrm{perturb}}^{(0)}  = -\frac{\nu}{N} \mathrm{Z} <0
    \label{Eq:KS-sync-perturb-LE}
\end{equation} where $\mathrm{Z}$ is again considered as an external forcing field. This Lyapunov exponent in fact corresponds to the perturbation along the sync-manifold (compare Eq.~(\ref{Eq:KS-global-sync-along-varEQ})). Note that this LE mainly depends on the collective behavior of the incoherent oscillators $\{ \phi_{i+N}(t) = s_m(t) | i=m=1,...,N \}$ (see Fig.~\ref{Fig:KS-LE-Poisson} (f)) via the summation term in Eq.~(\ref{Eq:KS-sync-perturb-LE}), i.e.,  the motion of the incoherent order parameter, and is much closer to zero than the transverse exponents in Eq.~(\ref{Eq:trans-LE-phase-global}). The CLV corresponding to $\Lambda_{\textrm{perturb}}^{(0)}$ has the form $\bold{v}_{\textrm{perturb}}^{(0)} = [v,...,v,v_1^{(\textrm{incoh})},...,v_N^{(\textrm{incoh})}]^\top   \in \mathbf{T}_{\bm{\phi}_{\textrm{ch}}(t)}(\mathbb{T}^{2N})$ where $\sum_{j=1}^{N} v_{j}^{\textrm{(incoh)}} \neq 0$. Hence, we conclude that all the Lyapunov modes (CLVs) in the synchronized population, both transverse and parallel to it, are stable, and therefore the synchronized manifold is invariant under the evolution of Eqs.~(\ref{Eq:KS-global-governing1}-\ref{Eq:KS-global-governing2}). Note that the Lyapunov exponents corresponding to the sync-population obtained here in Eq.~(\ref{Eq:trans-LE-phase-global}) and Eq.~(\ref{Eq:KS-sync-perturb-LE}) are consistent with previous results in Ref.~\onlinecite{abrams_chimera2016}. Therein, the authors considered the Jacobian matrix of the synchronized oscillator dynamics by treating the incoherent oscillators as external forcing functions, and then calculated the eigenvalues of the Jacobian matrix for the synchronized oscillators.

All the chimera states in a global two-population network, regardless of the parameters, i.e., also regardless of whether they are of the stationary or breathing type, have the $(N-1)$-fold degenerate $\Lambda_{\textrm{trans},\kappa}^{(0)}$ for $\kappa=2,...,N$ and $\Lambda_{\textrm{perturb}}^{(0)}$ since it is dictated by the symmetries of the global network topology and the perfectly synchronized oscillators. Thus, in Fig.~\ref{Fig:KS-LE-Poisson} (b), the same classes of the sync LEs for the breathing chimera state can be detected.

\subsubsection{\label{subsubsec:incoh-LE-KS-global} Incoherent Population: $\Lambda^{(\textrm{incoh})}_{\textrm{zero}}$ and $\Lambda^{(\textrm{incoh})}_{\rho}$}

Next, we turn to the  $(N-1)$-fold degenerate zero Lyapunov exponents $\Lambda_{\textrm{zero}}^{\textrm{(incoh)}}=0$ and the negative exponent $\Lambda_{\rho}^{(\textrm{incoh})} < 0$ of the stationary Poisson chimeras (Fig.~\ref{Fig:KS-LE-Poisson} (a)) that are associated with the incoherent oscillators. To better understand their origin, we consider the reduced dynamics according to the Watanabe-Strogatz transformation~\cite{WStheory,pikovsky_WS, mobius_strogatz}.
\begin{equation}
    \textrm{tan}\Bigg[ \frac{\phi_i^{(a)}-\Phi_a}{2} \Bigg] = \frac{1-\rho_a}{1+\rho_a}    \textrm{tan}\Bigg[ \frac{\psi_i^{(a)}-\Psi_a}{2} \Bigg]
\end{equation} where $a=1,2$ denotes the population index and $\psi_i^{(a)}$ are the constants of motion determined by the initial condition. This transformation leads to the 6-dimensional reduced set of equations~\cite{abrams_chimera2016}
\begin{flalign}
\frac{d \rho_a}{dt} &= \frac{1-\rho^2_a}{2}\textrm{Re}\bigg ( H_a e^{-i \Phi_a} \bigg)  \notag  \\ 
\frac{d \Psi_a}{dt} &= \frac{1-\rho^2_a}{2\rho_a}\textrm{Im}\bigg ( H_a e^{-i \Phi_a} \bigg) \label{Eq:WSequation}   \\
\frac{d \Phi_a}{dt} &= \frac{1+\rho^2_a}{2\rho_a}\textrm{Im}\bigg ( H_a e^{-i \Phi_a} \bigg) \notag
\end{flalign} for $a=1,2$. The mean-field forcing $H_a$ is given by
\begin{flalign}
H_1 &= \mu e^{-i(\alpha-\Phi_1)} \rho_1 \gamma_1 + \nu e^{-i(\alpha-\Phi_2)} \rho_2 \gamma_2 \notag \\
H_2 &= \mu e^{-i(\alpha-\Phi_2)} \rho_2 \gamma_2 + \nu e^{-i(\alpha-\Phi_1)} \rho_1 \gamma_1 \notag
\end{flalign} where $\gamma_a$ is defined by the same way in Eq.~(\ref{Eq:gamma-definition}) for each population. The 6-dimensional reduced dynamics in Eq.~(\ref{Eq:WSequation}) with the tangent space dynamics along the corresponding chimera reference trajectory ($\rho_1(t) =1$ and $\rho_2(t) <1$) is associated with six Lyapunov exponents which can be determined numerically. In Fig.~\ref{Fig:KS-LE-Poisson} (c-d) their values are shown versus the index for the same parameters which were used in the calculations of the full Lyaponov spectra depicted in  Fig.~\ref{Fig:KS-LE-Poisson} (a-b). The results give further insight on the LEs of the incoherent population: the incoherent WS reduced dynamics resides in an invariant subspace of the phase space of the incoherent population that is determined by the $N-3$ constants of motion, i.e., by the initial condition~\cite{WStheory} (here, $\textbf{PIC}$s and the uniform distribution of the constants of motion consistent with the Poisson submanifold), which yield $N-3$ neutral directions, i.e., $N-3$ zero LEs. In addition, there are two further zero exponents associated with the incoherent population that come from the two angular variables ($\Phi_2$, $\Psi_2$) in the reduced dynamics~\cite{pikovsky-nonlinear}. Hence, we obtain in total $N-1$ zero exponents. Apart from these zero LEs, there exists one negative LE that corresponds to the stable fixed point of the radial variable $\rho_2(t) \sim \rho_0 = \textrm{const.}$ whose value is determined by the parameter set. (Note that the remaining exponents in the WS reduced dynamics arise from the sync-group and the continuous time-shift symmetry.) Regarding the breathing chimera states, we find $N$-fold degenerate zero exponents in the incoherent population; an additional zero Lyapunov exponent results from the oscillating nature of the WS radial variable, i.e., the breathing motion of the order parameter of the incoherent population above the Hopf bifurcation~\cite{abrams_chimera2008, abrams_chimera2016}.

\subsubsection{\label{subsubsec:collective_modes_KSglobal} Collective Modes in Poisson Chimeras}

As a last step of our analysis of the dynamics of Poisson chimeras, we investigate whether some of the CLVs correspond to collective perturbations, or modes. Therefore, we calculate the time-averaged \textit{inverse participation ratio} (IPR) for various system sizes according to~\cite{ ipr, kevin}
\begin{equation}
    \textrm{IPR}^{(i)}(N) = \Bigg< \textrm{exp} \Bigg( \frac{1}{q-1} \textrm{log} \sum_{j=1}^{2N} \bigg| v^{(i)}_{j}(t) \bigg|^{2q} \Bigg) \Bigg>_t
    \label{Eq:ipr-definition}
\end{equation} where $q=2$ and $\textrm{IPR}^{(i)} \in [(2N)^{-1},1]$ and $v_{j}^{(i)}$ is the $j^{\textrm{th}}$ component of the CLV $\bold{v}^{(i)}  \in \mathbf{T}_{\bm{\phi}_{\textrm{ch}}(t)}(\mathbb{T}^{2N}) $ corresponding to a given exponent denoted by $\Lambda_i(N)$ defined in Eq.~(\ref{Eq:LE-definition}) for $i=1,...,2N$. By definition, $\textrm{IPR}^{(i)}(N)$ is close to $1$ if the given vector is well localized but close to $\frac{1}{2N}$ if the vector components spread out through all the oscillators. Therefore, a CLV is a collective mode if $\textrm{IPR}^{(i)}(N) \sim \frac{1}{N}$ as $N$ increases, whereas a CLV is localized when $\textrm{IPR}^{(i)}(N) \sim \textrm{const.}$ as $N$ increases~\cite{kevin, ipr}.

In Fig.~\ref{Fig:KS-LE-Poisson} (e-f), the numerically obtained IPRs of the CLVs corresponding to $\Lambda_{\textrm{perturb}}^{(0)}$ and $\Lambda_{\rho}^{\textrm{(incoh)}}$ of the stationary Poisson chimera are plotted versus the system size. The proportionality of $\textrm{IPR}(N) \sim \frac{1}{N}$ for large $N$ strongly suggests that the corresponding CLVs are indeed Lyapunov collective modes. As discussed above, these modes are related to the incoherent oscillators and affected by the incoherent order parameter motion. This observation is confirmed by our Lyapunov analysis, i.e., by measuring the localization of the covariant Lyapunov vector. We stress that these Lyapunov modes ($\Lambda_{\textrm{perturb}}^{(0)}$ and $\Lambda_{\rho}^{\textrm{incoh}}$) are collective (non-localized) throughout all the oscillators, and not restricted to the incoherent oscillator population.

\subsubsection{\label{subsubsec:incoh-LE-KS-global-nPIC} Non-Poisson Chimeras}

\begin{figure}[t!]
\includegraphics[width=1.0\linewidth]{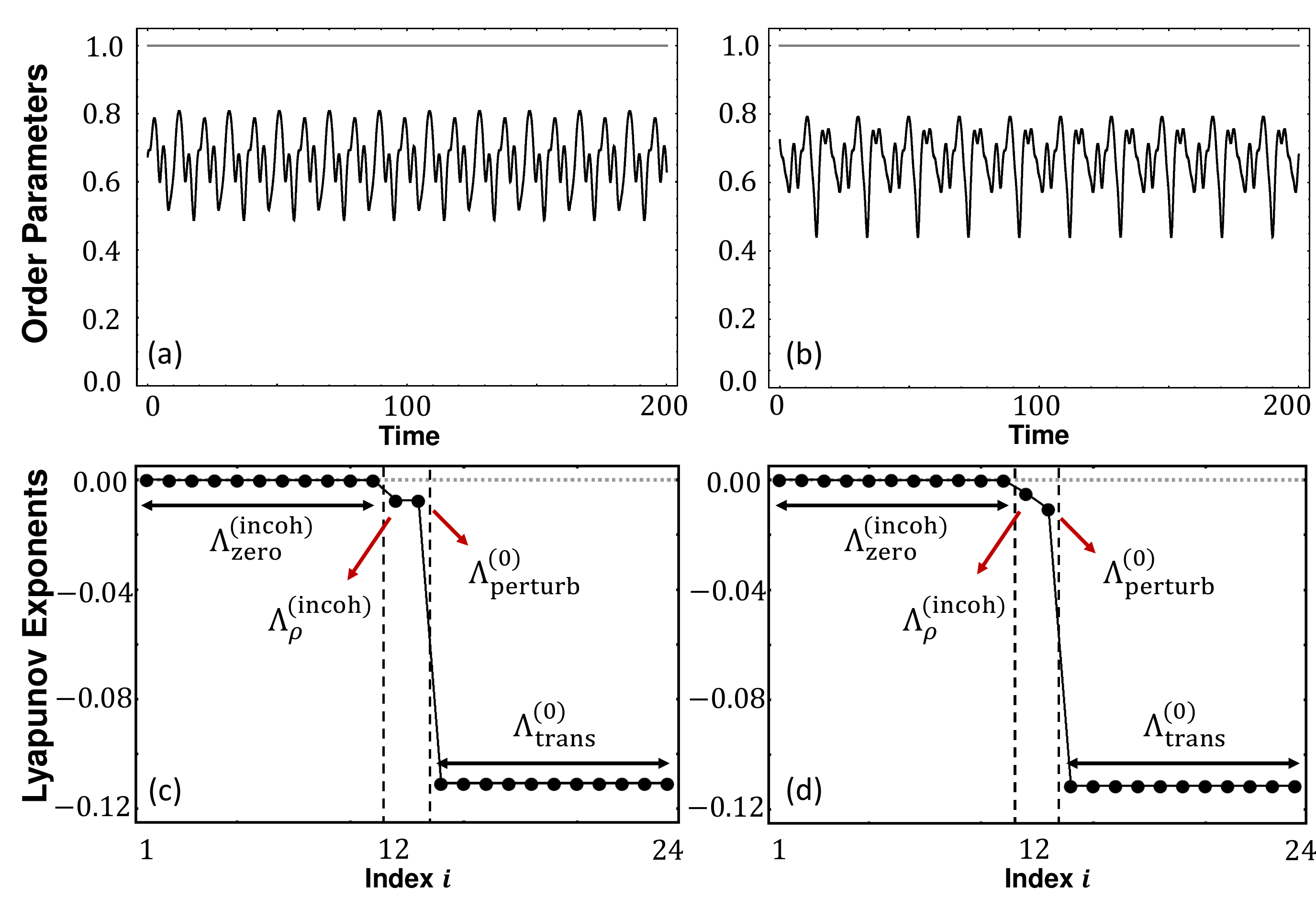}
\caption{(a-b) Temporal evolution of the magnitude of the Kuramoto order parameter obtained from non-Poisson chimera times series starting from different $ \textbf{n-PIC} $s after a time $ t \geq 10^6 $ for $ N=12 $ and $ A=0.2 $. (c-d) Lyapunov spectra corresponding to the dynamics of (a-b).} 
\label{Fig:KS-LE-Random}
\end{figure}

Finally, we turn to the Lyapunov exponents of the non-Poisson chimera trajectories that start from a given $\textbf{n-PIC}$. Two examples of the temporal evolution of magnitude of the order parameter of non-Poisson chimera trajectories that were obtained from different \textbf{n-PIC} but otherwise identical parameters in the governing equations are depicted in Fig.~\ref{Fig:KS-LE-Random} (a-b) together with the corresponding numerically determined Lyapunov spectra (c-d). In line with our discussion above in Sec.~\ref{subsec:random-ch}, non-Poisson chimera trajectories show different incoherent motions of the order parameter depending on a given \textbf{n-PIC}. In spite of this, since a non-Poisson chimera also lives on the two-population network, there are also $(N-1)$-fold degenerate $\Lambda_{\textrm{trans},\kappa}^{(0)}$ for $\kappa=2,...,N$ of the synchronized population given by Eq.~(\ref{Eq:trans-LE-phase-global}). Likewise, the numerical CLV analysis confirms these are indeed transverse to the sync-manifold as in Eq.~(\ref{Eq:trans-CLV-KS-global}). What is different from Poisson chimeras, particularly in the synchronized population, is that the LE arising from the perturbation along the sync-manifold ( Eq.~(\ref{Eq:KS-sync-perturb-LE})) takes a different value than in Poisson chimeras. This is because $\Lambda_{\textrm{perturb}}^{(0)}$ strongly depends on the motion of the incoherent oscillators through $\mathrm{Z}$ in Eq.~(\ref{Eq:KS-sync-perturb-LE}), which is determined by the initial condition. 

Concerning the LEs in the incoherent population, $(N-1)$-fold degenerate $\Lambda_{\textrm{zero}}^{(\textrm{incoh})}=0$ are also found from the WS reduced dynamics. However, since $\Lambda_{\rho}^{\textrm{(incoh)}}$ strongly depends on the constants of motion determined by the non-Poisson initial condition, it also attains a value different from that of a Poisson chimera trajectory, See Fig.~\ref{Fig:KS-LE-Random} (c-d).

\section{\label{sec:attracing-Poisson}Two ways to attracting Poisson chimera}

So far, many authors have observed that a small heterogeneity, e.g., nonidentical natural frequencies or noisy oscillators, makes the dynamics evolve towards at least a close neighborhood of the OA manifold and Poisson submanifold for the continuum limit and finite size system, respectively, and this stabilizing effect has been reported to be a generic consequence of the heterogeneity of the dynamics~\cite{pikovsky_WS,pikovsky-heteroWS2,pikovsky-heteroWS3,OA2,OA-attractive,Laing_hetero,Laing_hetero2}. In this section, we study two simple systems with identical oscillator populations that, according to the Lyapunov analysis, possess attracting Poisson chimeras. In the first system, we consider a nonlocal intra-population coupling, in the second one amplitude degrees of freedom of the oscillators, i.e., we employ Stuart-Landau amplitude oscillators rather than phase oscillators.

\subsection{\label{subsec:nonlocal} Topological variation: nonlocal intra-population network}

While previous studies on nonlocal intra-population networks focused on randomly but systematically constructed topologies and on chimera states in the continuum limit~\cite{Laing_nonlocal,timevarying_twogroup}, we consider here the simplest regular and finite-sized nonlocal network. This allows us to take advantage of the symmetry of the network. As depicted in Fig.~\ref{Fig:network}(b) the oscillators of each population are arranged on a ring. Compared to the globally coupled intra-population network, each oscillator has one intra-population connection less: it is not connected to the opposite oscillator. For this purpose, we only consider even numbers of the oscillators in each population here.  The adjacency matrix of this nonlocal intra-population but global inter-population network is defined as 

\begin{equation}
A=
\begin{pmatrix}
   \coolover{$N/2$}{ 0 & 1 & \cdots & 1  } &\vline & \coolover{$N/2$}{ 0 & 1 & \cdots & 1  } \\ 
    1 & 0 & \ddots & \vdots & \vline &    1 & 0 & \ddots& \vdots \\
    \vdots & \ddots & \ddots &1 & \vline &  \vdots & \ddots & \ddots &1 \\
    1 & \cdots & 1 & 0 & \vline & 1 & \cdots & 1 & 0 \\ \cline{1-9}
    0 & 1 & \cdots & 1 &\vline &  0 & 1 & \cdots & 1   \\ 
    1 & 0 & \ddots & \vdots & \vline &    1 & 0 & \ddots& \vdots \\
    \vdots & \ddots & \ddots &1 & \vline &    \vdots & \ddots & \ddots &1 \\
    1 & \cdots & 1 & 0 & \vline &   1 & \cdots & 1 & 0
    \end{pmatrix} 
    \label{Eq:nonlocal-adjacency-matrix}
\end{equation}
where the $i$-th oscillator is  disconnected to $(i+\frac{N}{2})$-th oscillator of the same population. 

\begin{figure}[t!]
\includegraphics[width=1.0\linewidth]{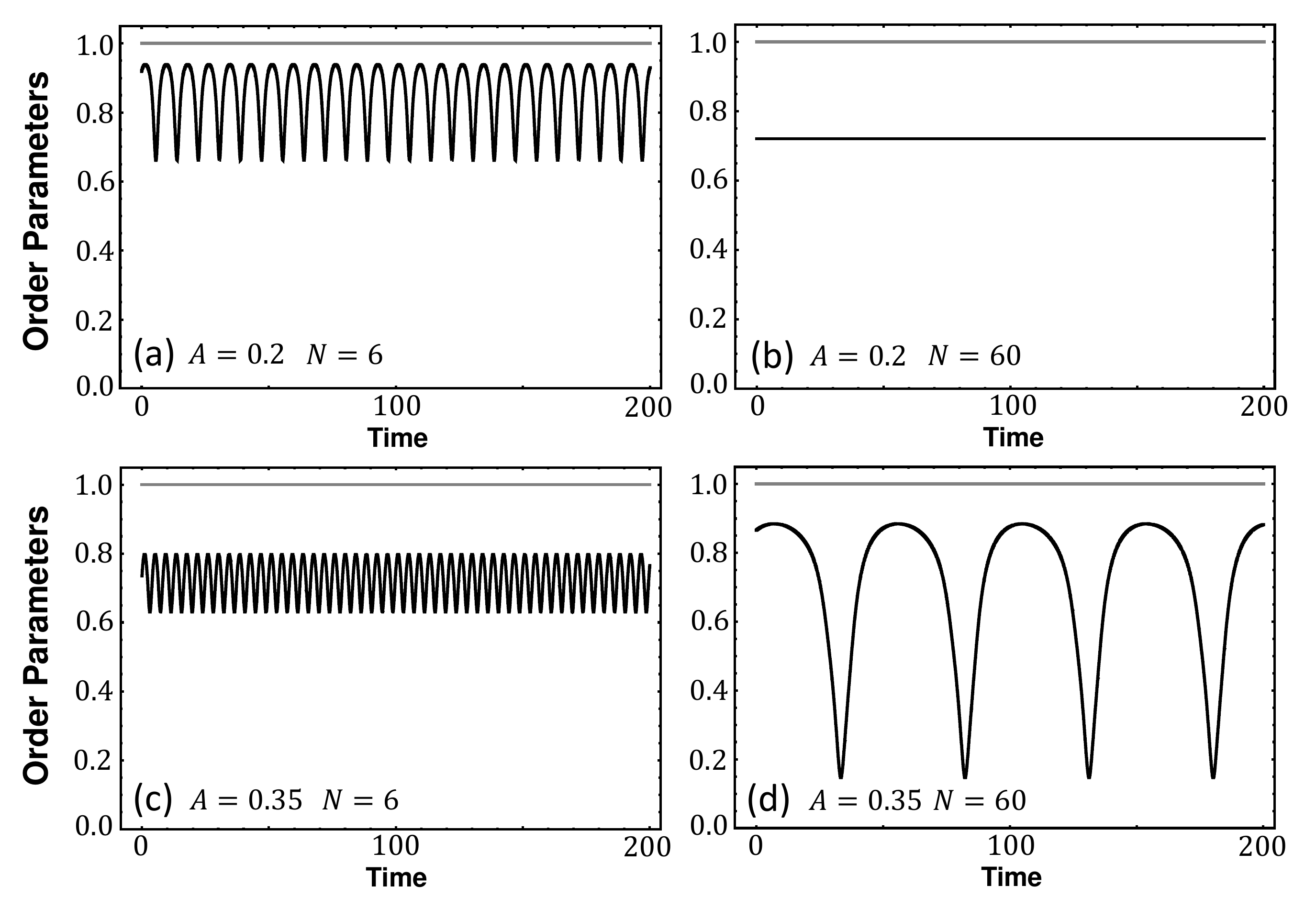}
\caption{The Kuramoto order parameters of the phase oscillators governed by the nonlocal intra-group coupling. (a,b) Chimera states with the parameter $A=0.2$ and $\beta=0.08$ corresponding to stationary chimeras for the system sizes $N=6$ and $N=60$, respectively. (c-d) Chimera sates with $A=0.35$ corresponding to the breathing chimera states. Gray line: synchronized group ($r(t)=1$),  black line: incoherent order parameter ($r(t)<1$).  } 
\label{Fig:Nonlocal-KS-order-parameter}
\end{figure}

The governing equations of the Kuramoto-Sakaguchi phase oscillators in the nonlocal intra-population topology are for the first population 
\begin{flalign}
\frac{d\phi_i}{dt} &=  -\frac{\mu}{N}\textrm{sin}\alpha+ \frac{\mu}{N}\sum_{j=1}^{N}A_{ij}\textrm{sin}(\phi_j-\phi_i-\alpha)\notag \\ &+\frac{\nu}{N}\sum_{j=1}^{N}\textrm{sin}(\phi_{j+N}-\phi_i-\alpha)
\label{Eq:KS-nonlocal-governing1}
\end{flalign}
for $i=1,...,N$, and
\begin{flalign}
\frac{d\phi_{i+N}}{dt} &=  -\frac{\mu}{N}\textrm{sin}\alpha+ \frac{\mu}{N}\sum_{j=1}^{N}A_{ij}\textrm{sin}(\phi_{j+N}-\phi_{i+N}-\alpha)\notag \\ &+\frac{\nu}{N}\sum_{j=1}^{N}\textrm{sin}(\phi_{j}-\phi_{i+N}-\alpha)
\label{Eq:KS-nonlocal-governing2}
\end{flalign}
for $i=1,...,N$ for the second one. $A_{ij}$ is the adjacency matrix that describes the nonlocal intra-population coupling of each population defined in Eq.~(\ref{Eq:nonlocal-adjacency-matrix}).

Although the nonlocally coupled system does not have a corresponding OA dynamics, we found that as long as we started from  $\textbf{PIC}$ $\in \mathrm{M}_{\textrm{Poisson}}$, chimera trajectories satisfy the dynamical characteristics of Poisson chimeras as defined in Sec.~\ref{subsec:poisson-ch}. For this nonlocal Poisson chimera state, the distribution of the incoherent phases remains in a close vicinity of the Poisson submanifold defined by Eq.~(\ref{Eq:poisson-kernel}) as the Poisson chimera distributions shown in Fig.~\ref{Fig:KS-small-size Poisson chimera} (c,g). Additionally, the nonlocal Poisson chimera also show the splay form of the instantaneous frequencies of the incoherent oscillators if $A=0.2$. In Fig.~\ref{Fig:Nonlocal-KS-order-parameter}, the simple motion of the magnitude of the order parameter of nonlocal stationary and breathing Poisson chimera states are depicted. For the parameter $A=0.2$, the magnitude of the order parameter has a practically constant value for large size chimeras (slightly different from the global topology), and the small-size chimera displays the clear periodic motion that arises from the splayed instantaneous velocities. For the parameter $A=0.35$, the order parameter of the large-size chimera state exhibits a main breathing motion as expected; however, the one of small-size chimeras does not show the main breathing motion superimposed by a secondary oscillation but rather it looks like that of the stationary Poisson chimera state.This might be interpreted as a hint that the nonlocality on the two-population network topology changes the Hopf bifurcation point for the small-size Poisson chimera as described for different non-complete networks in Ref.~\onlinecite{Laing_nonlocal}.

\subsection{Lyapunov analysis of Poisson chimeras in the nonlocal intra-population network\label{subsec:KS-nonlocal-LEfull}}

\subsubsection{\label{subsubsec:sync-LE-KS-nonlocal} Synchronized Population: $\Lambda_{\textrm{trans}}^{(0)}$ and $\Lambda_{\textrm{perturb}}^{(0)}$}

\begin{figure}[t!]
\includegraphics[width=1.0\linewidth]{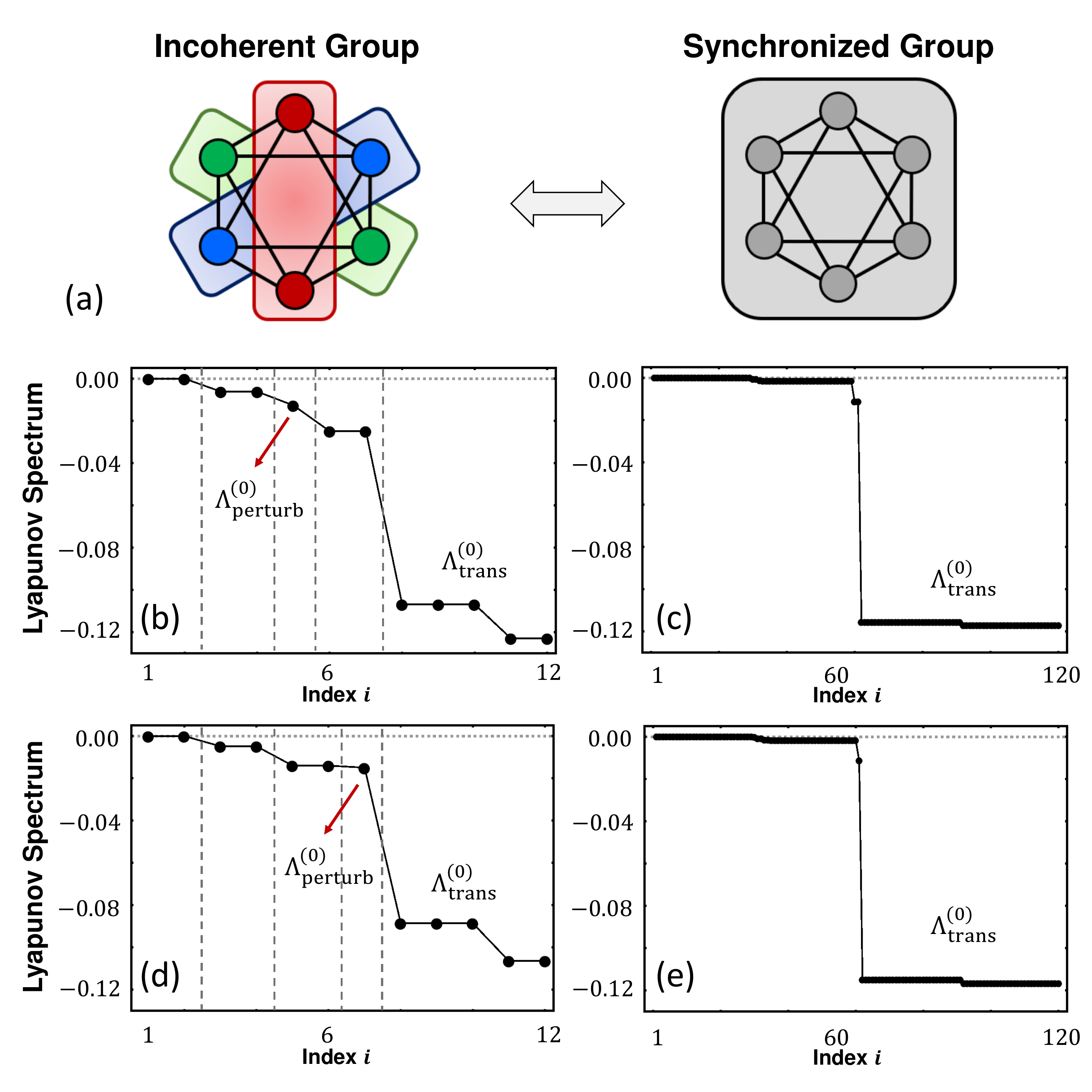}
\caption{ (a) Schematic drawing of the two-population oscillators metwork with nonlocal coupling for $N=6$. The same color in the incoherent group indicates that the oscillators marked by the same color are characterized by the same evolution dynamics. (b-c) The Lyapunov spectra for $A=0.2$ with $N=6$ and $N=60$, respectively. (d-e) The Lyapunov spectra for $A=0.35$ with $N=6$ and $N=60$, respectively. } 
\label{Fig:LE-nonlocal-phase}
\end{figure}

For the Poisson chimeras on the nonlocal topology, the Lyapunov spectrum of the nonlocal Poisson chimeras is qualitatively different from the one of the global Poisson chimeras, as can be seen in Fig.~\ref{Fig:LE-nonlocal-phase}. There are $N-1$ transverse Lyapunov exponents consisting of two different values. This splitting of the values of $\Lambda_{\textrm{trans}}^{(0)}$ is due to the fact that the transversal variational equations include two different eigenvalues of the adjacency matrix corresponding to the same synchronized cluster according to the nonlocal network symmetry (compare Eqs.~(\ref{Eq:transverse-nonlocal-phase}-\ref{Eq:KS-nonlocal-trans-LE})). The analytical approximate expressions of the $N-1$ transverse Lyapunov exponents to the sync-manifold are
\begin{equation}
   \Lambda_{\textrm{trans},\kappa}^{(0)} =\begin{dcases}
    -\frac{\mu}{N}(N-2)\textrm{cos}\alpha -\frac{\nu}{N}\mathrm{Z} <0, & \kappa =  2,...,N/2+1 \\ \\
     - \mu\textrm{cos}\alpha -\frac{\nu}{N}\mathrm{Z}<0, & \kappa=N/2+2,...,N
  \end{dcases}
  \label{Eq:Nonlocal-KS-transverseLE}
\end{equation} provided that $\mathrm{Z}$ is treated as an external forcing field. The numerical calculation of the CLVs confirms that the LEs in Eq.~(\ref{Eq:Nonlocal-KS-transverseLE}) are indeed transverse to the sync-manifold since the corresponding CLVs have the form $\bold{v}_{\kappa}^{(0)} = [ v_{\kappa 1}^{(\textrm{trans})},...,v_{\kappa N}^{(\textrm{trans})},0,...,0]^\top \in \mathbf{T}_{\bm{\phi}_{\textrm{ch}}(t)}(\mathbb{T}^{2N}) $  while $\sum_{i=1}^{N}v_{\kappa i}^{(\textrm{trans})}=0$ for $\kappa=2,...,N$. Note that as $N$ increases, the gap between the transverse Lyapunov exponents in Eq.~(\ref{Eq:Nonlocal-KS-transverseLE}) is decreasing, and the numerical results in Fig.~\ref{Fig:LE-nonlocal-phase} reflect this fact.

Also, as can be seen in Fig.~\ref{Fig:LE-nonlocal-phase} there is another LE of the synchronized population, which arises from a perturbation along the sync-manifold. This perturbation brings forth the very negative exponent $\Lambda_{\textrm{perturb}}^{(0)} = -\frac{\nu}{N}\mathrm{Z}<0$ that strongly depends on the motion of the incoherent oscillators. Hence, we conclude that in the nonolocal intra-population topology the synchronized population of Poisson chimera states is also stable in both the directions transverse and  parallel to the sync-manifold. 

\subsubsection{\label{subsubsec:incoh-LE-KS-nonlocal} Incoherent Population : Paris of Two Near-degenerate Lyapunov Exponents }

Next, we focus on the Lyapunov exponents corresponding to the incoherent oscillators. Although we cannot apply directly the Watanabe-Strogatz reduction ( Eq.~(\ref{Eq:WSequation})) in case of the nonlocally coupled oscillators, the classification of the incoherent LEs can be addressed as follows.  The quotient dynamics for the incoherent population in Eq.~(\ref{Eq:quotient_nonlocal-phase-incoherent}) contains discrete symmetries due to the topology of the nonlocal network (see Fig.~\ref{Fig:LE-nonlocal-phase} (a)). Since each oscillator is disconnected only from the opposite one, two oscillators $s_m(t)$ and $s_{m+N/2}(t)$ are characterized by the same evolution equation. It is also known that such discrete symmetries cause  near-degeneracy in the Lyapunov spectrum~\cite{pikovsky_LE}. Thus, $N/2$ pairs of two nearly degenerate exponents occur in the incoherent population (see Fig.~\ref{Fig:LE-nonlocal-phase} (b,d)). Therefore, unlike the globally coupled Poisson chimeras, which are neutrally stable, the incoherent population of nonlocal Poisson chimeras is stable, as suggested by the fact that all pairs of the incoherent Lyapunov exponents are definitely negative, except for the two zero  exponents which are connected to the continuous symmetries: the phase shift ($\bold{v}_{\textrm{ps}} = (\delta \phi_0,...,\delta \phi_0)^\top$ where $| \delta \phi_0 | \ll 1$) and the time shift ($\bold{v}_{\textrm{ts}} \propto \dot{\bold{\phi}}_{\textrm{ch}}=\bold{f}(\bold{\phi_{\textrm{ch}}})$), respectively, which in fact do not affect the stability of the trajectory~\cite{kevin}. For large-size Poisson chimeras in Fig.~\ref{Fig:LE-nonlocal-phase} (c,e), the near-degenerate pairs in the incoherent population are getting closer and closer to one another, until eventually, due to the nonlocal network symmetry, they tend to form two different continuous distributions, one of which consists of obviously negative LEs, whereas the other one consists of two (or some) zero and very slightly negative LEs, corresponding to slow but stable Lyapunov exponents (within our numerical ability). 

On the other hand, we can also think of this attractiveness of nonlocal Poisson chimeras due to the heterogeneity of the system. Our nonlocal topology is generated by the least change from the global topology, and hence if we make global the summation term in Eq.~(\ref{Eq:quotient_nonlocal-phase-incoherent}), then the disconnecting term due to the nonlocal topology between the two oscillators $s_m$ and $s_{m+N/2}$ should be included in the uncoupled term outside the summation and Eq.~(\ref{Eq:quotient_nonlocal-phase-incoherent}) becomes
\begin{flalign}
\frac{d s_m}{dt} = \tilde{\omega}_m(t) + \nu\textrm{sin}(s_0-s_m-\alpha) + \frac{\mu}{N}\sum_{m'=1}^{N}\textrm{sin}(s_{m'}-s_m-\alpha)
\label{Eq:hetero-nonloca}
\end{flalign}
with $\tilde{\omega}_m(t)=-\frac{\mu}{N}\textrm{sin}(s_{m+N/2}-s_m-\alpha) \sim \mathcal{O}(N^{-1})$. Thus we can interpret $\tilde{\omega}_m(t)$ as a small heterogeneity for the globally coupled incoherent oscillator population. Such a heterogeneity is known to confine the chimeras in a vicinity of the Poisson submanifold.~\cite{OA2, OA-attractive,pikovsky-heteroWS,pikovsky-heteroWS2,pikovsky-heteroWS3}

\begin{figure}[t!]
\includegraphics[width=1.0\linewidth]{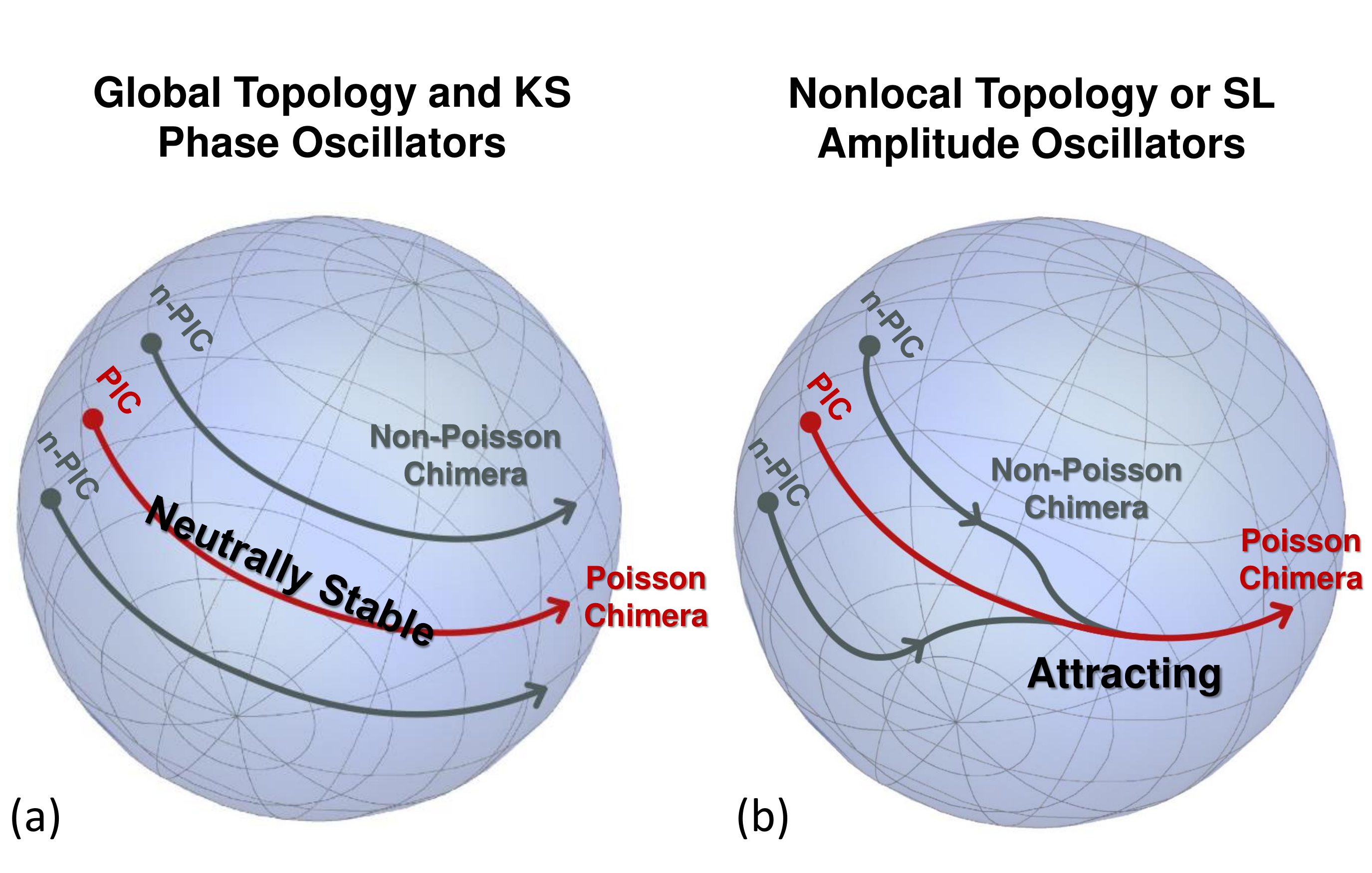}
\caption{Schematic representation of chimera trajectories in the invariant manifold $\mathrm{M}_{\textrm{incoh}}$ (sphere) and Poisson submanifold $\mathrm{M}_{\textrm{Poisson}} \subset \mathrm{M}_{\textrm{incoh}}$ (red curve). Each line schematically represents a trajectory of a chimera state from a given initial condition. The arrow indicates the time flow in the incoherent phase space. (a) For the Kuramoto-Sakaguchi phase oscillators on the global intra-group coupling; the incoherent trajectories of the chimeras dwell in the neutrally stable manifold $\mathrm{M}_{\textrm{incoh}}$. The Poisson chimera trajectories reside in the invariant and also neutrally stable Poisson submanifold only if the trajectory starts from $\textbf{PIC}$; the non-Poisson chimera from $\textbf{n-PIC}$ dwells in the manifold outside the Poisson submanifold. Thus, the non-Poisson chimeras exhibit various incoherent motion according to the given $\textbf{n-PIC}$. (b) Attracting Poisson chimeras for the nonlocal intra-group topology or Stuart-Landau oscillators. The trajectories starting even from $\textbf{n-PIC}$ eventually settle down on or close to the Poisson trajectory. } 
\label{Fig:Schematic-Manifold}
\end{figure}

\subsection{\label{subsec:SL-amplitude}Dynamical variation: Stuart-Landau oscillators}

As the second way to obtain attracting Poisson chimeras, we consider Stuart-Landau (SL) planar oscillators. This two-population network of SL oscillators has been studied recently in the continuum limit~\cite{Laing_SL2010,Laing_SL2019}, in which attracting chimeras states have been reported. Here, we consider the finite-sized ensemble and give a full Lyapunov stability analysis which gives further evidence that amplitude DOFs render Poisson chimeras attracting. The amplitude degrees of freedom introduce a small heterogeneity, which is, however, this time self-organized~\cite{Laing_SL2010,Laing_hetero2}.

In an ensemble of Stuart-Landau (SL) oscillators, each oscillator has a phase $\phi_i(t) \in [-\pi,\pi)$ and an amplitude $r_i(t) \in (0,\infty)$ variable. The governing equations are
\begin{flalign}
\frac{d r_i}{dt} &= \epsilon^{-1}(1-r^2_i)r_i + \frac{\mu}{N}\sum_{j=1}^{N}r_j\textrm{cos}(\phi_j-\phi_i-\alpha)\notag \\ &+\frac{\nu}{N}\sum_{j=1}^{N}r_{j+N}\textrm{cos}(\phi_{j+N}-\phi_i-\alpha)
\label{Eq:govern_globalamplitude}
\end{flalign}
for $i=1,...,N$, which depicts the evolution of the amplitude variables of the oscillators in the first oscillator population, and
\begin{flalign}
\frac{d \phi_i}{dt} &= \omega - \sigma r^2_i +\frac{\mu}{N}\sum_{j=1}^{N}\frac{r_j}{r_i} \textrm{sin}(\phi_j-\phi_i-\alpha)\notag \\ &+\frac{\nu}{N}\sum_{j=1}^{N}\frac{r_{j+N}}{r_i}\textrm{sin}(\phi_{j+N}-\phi_i-\alpha)
\label{Eq:govern_globalphase}
\end{flalign}
for $i=1,...,N$, describing the phase dynamics of the SL oscillators in the same population. The governing equations for the second population can also be easily obtained in the same way. In our further study, we fix some parameters: $\sigma=0.2$ and $\omega=0$. Notice that as $\epsilon \rightarrow 0$, the system approaches the evolution equations~(\ref{Eq:KS-global-governing1}-\ref{Eq:KS-global-governing2}) of the phase-only oscillators whose amplitude $r_i \rightarrow 1$ for all $i=1,...,2N$~\cite{Laing_SL2010}.

To study Poisson chimeras of the SL ensemble, we start from the $\textbf{PIC}$ on the phase variables in Eq.~(\ref{Eq:govern_globalphase}) in one population, and set the phases of the second population to the same value and all the initial amplitudes in Eq.~(\ref{Eq:govern_globalamplitude}) to $r_i(0)=1$ for $i=1,...,2N$ (Note that the definition of Poisson chimeras involves only the phase DOFs). The states evolving from such a $\textbf{PIC}$ satisfy all the dynamical properties in the definition of Poisson chimeras for the phase DOFs: one population remains perfectly synchronized, the incoherent phase distribution remains in the Poisson kernel, and finally large- and small-size behavior emerges according to the system size. In particular, the stationary chimera states show the splay form of the instantaneous incoherent frequencies that yield the periodic order parameter for the small-size stationary chimeras. Regarding the amplitude variables, all synchronized oscillators have an amplitude $r_i(t)=1$ for $i=1,...,N$ and the amplitudes of the oscillators in the other population show some distribution with the degree of variation depending on the parameter $\epsilon$.

For the SL oscillators, the coupling strength $\epsilon$ acts as a bifurcation parameter. For weak coupling strength, i.e., sufficiently small $\epsilon$ (here, we use $\epsilon=0.01$) the dynamics are close to the phase-reduced behavior. Hence, the evolution of the order parameter is very close to the one depicted in Fig.~\ref{Fig:KS-Poisson-chimera-orderparameter} for the phase reduced system; $r_{\textrm{incoh}}(t)$ is stationary for $A=0.2$ and exhibits a breathing motion for $A=0.35$. However, when increasing $\epsilon$ at constant $A=0.2$ the stationary chimera undergoes eventually a Hopf bifurcation, giving rise to breathing chimeras, which are observed, e.g., for $\epsilon=0.15$, which is in line with findings reported in Ref.~\onlinecite{Laing_SL2010}.

\begin{figure*}[t!]
\includegraphics[width=1.0\textwidth]{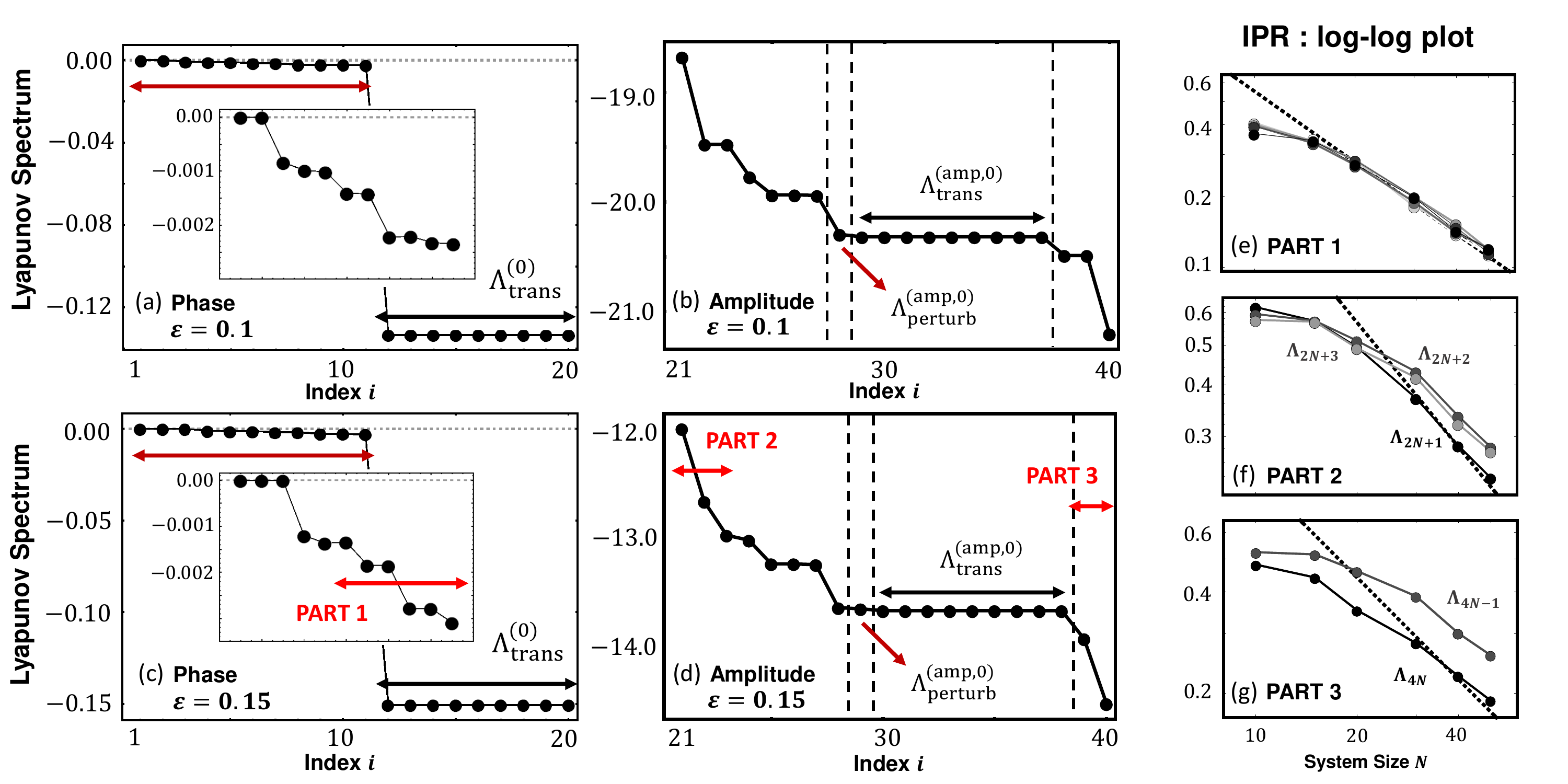}
\caption{Lyapunov exponents vs index of the strongly coupled SL oscillators with global intra-group coupling for $A=0.2$ and (a-b) $\epsilon=0.1$ (stationary) and (c-d) $\epsilon=0.1$ (breathing). (a,c) Lyapunov exponents of phase DOFs. The Insets  show a magnification of the Lyapunov exponents corresponding to the incoherent phase DOFs. (b,d) Lyapunov exponents corresponding to amplitude DOFs. (e-g) IPR versus system size $N$ for $\epsilon=0.15$ and Lyapunov modes corresponding corresponding to the exponents in PART 1 (e), PART 2 (f) and part 3 (g). The black dashed guidelines indicate $\sim 1/N$.} 
\label{Fig:strongly-SLE}
\end{figure*}

\subsection{\label{subsubsec:SL-global} Lyapunov analysis on Poisson chimeras of Stuart-Landau oscillators}

To study the Lyapunov exponents numerically, we exploit the real-valued coordinates of each Stuart-Landau oscillator~\cite{kevin}. The variables of an SL oscillator can be represented by
\begin{equation}
    r_k(t)e^{i\phi_k(t)} = \frac{1}{\sqrt{2}}\big( a_k(t) + i b_k(t)\big)
    \label{Eq:real-value-coordi}
\end{equation} for $k=1,...,2N$ where $a_k$ and $b_k$ are real-valued functions of time. Thus, the perturbation vectors in the tangent space are written in the form $\bold{v}^{(i)} = (a_1,...,,a_N,a_{N+1},...,a_{2N},b_{1},...,b_N,b_{N+1},...,b_{2N})^\top \in \mathbf{T}_{\bold{x}_{\textrm{ch}}(t)}(\mathbb{R}^{4N}) $. This coordinate transformation is a unitary transformation; hence, it can uphold the information on Lyapunov exponents.

\subsubsection{\label{subsubsec:sync-LE-SL-amplitude} Amplitude Degrees of Freedom}

In Fig.~\ref{Fig:strongly-SLE}, the numerically obtained Lyapunov spectra of chimera states for strong ($\epsilon=0.1$ (a,b) and $0.15$ (c,d)) coupling are displayed. The spectra are composed of two parts, which correspond to the phase and amplitude degrees of freedom, respectively. The former are shown in the left column, the latter in the middle one.

\begin{figure}[t!]
\includegraphics[width=1.0\linewidth]{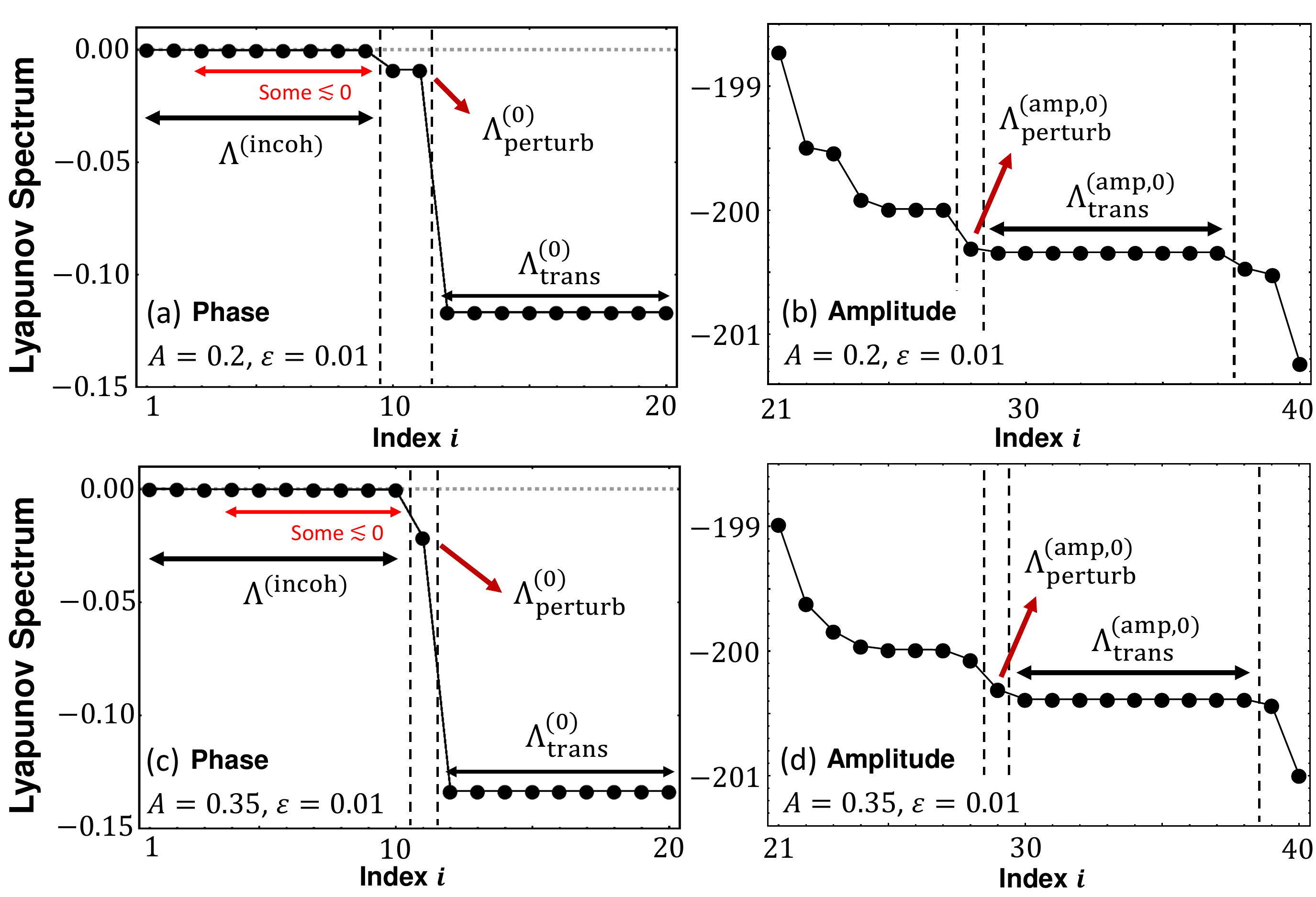}
\caption{Full Lyapunov spectra of the Stuart-Landau oscillators for weak coupling $\epsilon=0.01$ and (a,b) stationary Poisson chimeras with $A=0.2$ and breathing Poisson chimeras with $A=0.35$. Left column: phase DOFs, right column amplitude DOFs.  } 
\label{Fig:LE-global-SLE}
\end{figure}

First, consider the amplitude DOFs of the synchronized oscillators. They have $(N-1)$-fold degenerate strongly negative Lyapunov exponents, which are transverse to the sync-manifold. The approximate values of these Lyapunov exponents are
\begin{equation}
    \Lambda_{\textrm{trans},\kappa}^{(\textrm{amp},0)} \approx \epsilon^{-1}(1-3R^2_0)<0 \label{Eq:SL-global-Amp-LE-trans}
\end{equation} for $\kappa=2,...,N$ (see Eq.~(\ref{Eq:trans-varial-amplitude})). The numerically obtained CLVs confirm that these Lyapunov exponents are indeed transverse to the sync-manifold as they have the following form
\begin{flalign}
    \bold{v}_{\kappa}^{(\textrm{amp},0)}&=\big(a_{\kappa 1}^{(\textrm{amp},0)},...,a_{\kappa N}^{(\textrm{amp},0)},0,...,0, \notag \\
    &b_{\kappa 1}^{(\textrm{amp},0)},...,b_{\kappa N}^{(\textrm{amp},0)},0,...,0 \big)^\top \in \mathbf{T}_{\bold{x}_{\textrm{ch}}(t)}(\mathbb{R}^{4N}) \notag
\end{flalign} where $\sum_{i=1}^{N}a_{\kappa i}^{(\textrm{amp},0)}=\sum_{i=1}^{N}b_{\kappa i}^{(\textrm{amp},0)}=0$ for $\kappa=2,...,N$. In Fig.~\ref{Fig:strongly-SLE} (b,d), we observe another negative exponent in the synchronized population of the amplitude DOFs caused by the perturbation along the sync-manifold as in Eqs.~(\ref{Eq:SL-global-syncalong-VarEQ-AMP} - \ref{Eq:SL-global-syncalong-AMP}). For the analytical value of it one obtains
\begin{equation}
    \Lambda_{\textrm{perturb}}^{(\textrm{amp},0)} \approx \epsilon^{-1}(1-3R^2_0) +\mu \textrm{cos}\alpha  <0
\end{equation} which is a slightly greater Lyapunov exponent than the transverse ones $\Lambda_{\textrm{trans}}^{(\textrm{amp},0)} \lesssim \Lambda_{\textrm{perturb}}^{(\textrm{amp},0)}$, in line with the numerical observations in Fig.~\ref{Fig:strongly-SLE}. The numerical CLV analysis reveals that this LE has the form $\bold{v}_{\textrm{perturb}}^{(\textrm{amp},0)}=(a,...,a,a_1^{(\textrm{inc})}...,a_N^{(\textrm{inc})},b,...,b,b_1^{(\textrm{inc})},...,b_N^{(\textrm{inc})})^\top  \in \mathbf{T}_{\bold{x}_{\textrm{ch}}(t)}(\mathbb{R}^{4N})$ where $a,b$ $\in \mathbb{R}$ are constant and  $\sum_{j=1}^{N}a^{(\textrm{inc})}_j \neq 0$ and $\sum_{j=1}^{N}b^{(\textrm{inc})}_j \neq 0$. Hence, we conclude that there is no perturbation direction in the amplitude DOFs, which corresponds to an unstable direction of the synchronized manifold, i.e., the sync-manifold remains invariant under the dynamics since for the sync-population in the amplitude DOFs the CLV modes both transverse and parallel to the sync-manifold are stable. For the other Lyapunov exponents in the amplitude DOFs, we guess that these stable Lyapunov exponents of the amplitude DOFs are linked to the incoherent oscillators through their quotient dynamics in Eq.~(\ref{Eq:incoh-amp-quotient}). Therefore, all the amplitude Lyapunov exponents are strongly negative, and the Poisson chimeras are strongly attracting in all the amplitude DOFs. Note that the amplitude DOFs of the SL ensemble for the weak coupling ($\epsilon=0.01$), depicted in Fig.~\ref{Fig:LE-global-SLE}, show the same behavior.

\subsubsection{\label{subsubsec:sync-LE-SL-phase} Phase Degrees of Freedom}

In the phase degrees of freedom, the synchronized oscillators also have the $(N-1)$-fold degenerate transverse Lyapunov exponents in Fig.~\ref{Fig:LE-global-SLE} and \ref{Fig:strongly-SLE}, whose analytical approximate expressions are
\begin{flalign}
\Lambda_{\textrm{trans},\kappa}^{(0)} &=  -\mu \textrm{cos}\alpha -\frac{\nu}{N} \sum_{m'=1}^{N}\frac{R_{m'}}{R_0} \textrm{cos}(s_{m'}-s_0-\alpha) \notag \\ &= -\mu \textrm{cos}\alpha -\frac{\nu}{N}\tilde{\mathrm{Z}}  < 0
\end{flalign} for $\kappa=2,...,N$ where $\tilde{\mathrm{Z}}=\sum_{m'=1}^{N}\frac{R_{m'}}{R_0} \textrm{cos}(s_{m'}-s_0-\alpha)$ should be considered as an external forcing field. In addition, the LE in the sync group coming from a perturbation along the sync-manifold has the value of $\Lambda_{\textrm{perturb}}^{(0)} =- \frac{\nu}{N}\tilde{\mathrm{Z}} <0$ and is expected to be found in the synchronized phase DOFs. The numerical CLV analysis also confirms that $\Lambda_{\textrm{trans},\kappa}^{(0)}$ and $\Lambda_{\textrm{perturb}}^{(0)}$ associated with the synchronized population are indeed transverse and parallel to the sync-manifold, respectively.

What makes Poisson chimeras of SL oscillators attractive are the incoherent LEs $\Lambda^{(\textrm{incoh})}$ in Fig.~\ref{Fig:strongly-SLE} (see inset) and Fig.~\ref{Fig:LE-global-SLE} (a,c). In an appropriate rotating reference frame, the quotient governing equations for the incoherent phase DOFs in Eq.~(\ref{Eq:SLE-phase-quotient}) are the same as for the phase-only oscillators in Eqs.~(\ref{Eq:KS-sync-cluster}-\ref{Eq:KS-incoherent-clusters}) except for the amplitude variables that can be considered as a small self-organized heterogeneity $\tilde{\Omega}_m(t)$ in the phase governing equations, Eq.~(\ref{Eq:SLE-phase-quotient})~\cite{Laing_SL2010}. For the strongly coupled systems with $\epsilon=0.1$ and $0.15$ as in Fig.~\ref{Fig:strongly-SLE} (a,c), there are clearly negative Lyapunov exponents in the incoherent phase DOFs. For the stationary chimera ($\epsilon=0.1$), we have, in addition, two zero exponents, for the breathing chimera ($\epsilon=0.15$), besides the negative exponents, there are three zero exponents, one of which arises from the oscillatory nature of the breathing chimeras. The stable Lyapunov exponents arise due to the amplitude variables in the phase governing equations which present a heterogeneity that, in turn, renders the chimeras attractive~\cite{OA2,Laing_hetero,Laing_hetero2,Laing_SL2010,OA-attractive,pikovsky-heteroWS,pikovsky-heteroWS2,pikovsky-heteroWS3}. For the weak coupling case $\epsilon=0.01$ in Fig.~\ref{Fig:LE-global-SLE} (a,c), the amplitude fluctuations are not that strong ($R_m \approx R_0=1$ for $m=1,...,N$) and as a result, Eq.~(\ref{Eq:SLE-phase-quotient}) can be approximated by Eqs.~(\ref{Eq:KS-sync-cluster}-\ref{Eq:KS-incoherent-clusters}) like the phase-reduced model, and the Poisson chimeras and their Lyapunov exponents follow patterns similar to the ones obtained for the KS oscillators (see Fig.~\ref{Fig:LE-global-SLE} (a,c) compared to Fig.~\ref{Fig:KS-LE-Poisson}). However, there is still a heterogeneity of the amplitude DOFs in the phase governing equations, and therefore we can expect the LEs to be still slightly negative (stable Lyapunov exponents) in the incoherent phase DOFs. Even for cases where these exponents are very close to zero in our numerical ability, compared to the KS phase-only LEs in Fig.~\ref{Fig:KS-LE-Poisson}, they are slightly decreasing to negative values in the index order which does not occur in the KS phase-only system. Hence, we tentatively conclude that also for weak coupling the stationary chimeras have only two zero LEs, all other exponents are weakly stable. Hence, in all cases, the Poisson chimeras are either at least weakly stable or clearly attracting, compared to the KS phase-only Poisson chimera states because the amplitude variables introduce a self-organized heterogeneity in the phase governing equations. As a consequence, even if starting from $\textbf{n-PIC}$, the chimera trajectories eventually approach the Poisson submanifold, as we could confirm with numerical simulations. 

More than this, we also exploited weakly coupled ($\epsilon=0.01$) SL oscillators in the nonlocal intra-population network in order to see whether the Poisson chimeras are also attracting or not. The detailed results on the Lyapunov analysis are compiled in Appendix.~\ref{append:concurrent}. In this case, the nonlocal topology leads to a stronger negative Lyapunov exponents than the globally coupled SL oscillators rather similar to the phase-only system in Fig.~\ref{Fig:LE-nonlocal-phase}. 
Therefore, the simultaneous perturbations also cause the Poisson chimeras to evolve towards a close neighbourhood of the Poisson submanifold, i.e. the Poisson chimera trajectory.

Finally, we also investigate whether the system has a Lyapunov collective mode or not by numerically evaluating the IPR function defined in Eq.~(\ref{Eq:ipr-definition}), especially for the case of the breathing chimera $\epsilon=0.15$. As can be seen in Fig.~\ref{Fig:strongly-SLE} (e-g), for at least six Lyapunov modes the IPR shows the tendency to decrease according to  $\textrm{IPR}^{(i)}(N) \sim \frac{1}{N}$ as the system size $N$ increases. This strongly suggests that these modes, which correspond to the negative exponents in PART 1 in Fig.~\ref{Fig:strongly-SLE}, are collective modes (Note that the stationary chimera $\epsilon=0.1$ shows the same collective modes, not shown here). In PART 2 and PART 3 in Fig.~\ref{Fig:strongly-SLE} for the amplitude DOFs, within our numerically tractable system sizes, at least one Lyapunov mode satisfies the inverse-proportional behavior of the IPR as a function of the system size in each PART, respectively. Consequently, these Lyapunov modes, $\Lambda_{2N+1}$ in PART 2 and $\Lambda_{4N}$ in PART 3 are not localized but affect all the oscillators collectively (not restricted only on the incoherent group, but spread out over all the oscillators) and are strongly related to their collective motion in the state space, i.e., they are also Lyapunov collective modes.

\section{\label{sec:conclusion} Conclusion}

In this work, we have dealt with chimera states in two-population networks of identical oscillators. For the identical Kuramoto-Sakaguchi phase oscillators, the order parameter dynamics of the incoherent oscillator population strongly depends on the initial condition and the population size~\cite{abrams_chimera2016,pikovsky_WS}. Once chimeras started from a special initial condition where all the initial phases of one population are in the Poisson kernel, i.e., the Poisson submanifold~\cite{mobius_strogatz,Laing_hetero2,OA,WStheory}, the phases remain in the Poisson kernel for all times, and we called this chimera a Poisson chimera. Poisson chimeras show a rather simple motion of the incoherent oscillator population that is virtually indistinguishable from the continuum limit OA solution for sufficiently large population sizes~\cite{abrams_chimera2008}. 
In contrast, the incoherent motion of a Poisson chimera with a small population size is drastically different from the simple OA dynamics~\cite{abrams_chimera2016}. This difference is not due to finite-size fluctuations, but has a  deterministic origin: 
The magnitude of the order parameter of the incoherent oscillator population shows not only the main motion close to the OA dyanmics but also a superimposed secondary oscillation along the main motion. We demonstrated that this superposed oscillation is a consequence of the fact that the instantaneous frequencies of stationary Poisson chimeras exhibit a splay-form. Furthermore, the splayed distribution of the instantaneous frequencies bring about that the period of the superposed oscillation tends to zero with increasing $N$ while 
the consideration of the WS global variables revealed how the amplitude of the secondary oscillation disappears with increasing $N$~\cite{pikovsky_WS,abrams_chimera2016,pikovsky-heteroWS}.
Consequently, our investigations have revealed that and how the order parameter changes continuously from small-size chimeras to large-size chimeras up to the continuum limit, eventually showing the same dynamics as the OA dynamics in the continuum boundary.

In contrast to such Poisson chimeras, the chimeras initialized outside the Poisson submanifold, called in this work non-Poisson chimeras, do not show such a simple order parameter dynamics, regardless of the system size, nor splay-formed instantaneous frequencies of the stationary chimeras, nor does the phase distribution stay in the Poisson kernel. Rather, they show complicated fluctuations along the main motion close to the OA dynamics. This complex, superposed trajectory exists for  stationary as well as breathing chimeras,  it does not disappear for the large population sizes and in the long time limit, and it depends on the particular initial condition outside the Poisson submanifold, i.e., a set of nonuniform constants of motion~\cite{pikovsky_WS,pikovsky-heteroWS,pikovsky-heteroWS2}. 

In our numerical Lyapunov analysis and also in other previous results~\cite{WStheory,pikovsky-nonlinear}, the stationary chimera states in two-population network with global intra- and inter-population coupling topology, whether it is a Poisson or non-Poisson chimera, are neutrally stable in $N-1$ directions. Note that the other negative LE corresponds to the degree of the coherence, i.e., the global WS radial variable. Based on the WS theory, the neutral stability mainly originates from the constants of motion of the system.
Any particular parameter set determines the value of the OA radial variable in the continuum limit. The phase DOFs of the neutrally stable Poisson chimeras are dictated by the Poisson initial condition, i.e.,  uniform constants of motion, and remain in the Poisson kernel~\cite{mobius_strogatz}. 

In contrast, the initial conditions for non-Poisson chimeras correspond to a non-uniform set of constants of motion that cause the different irregular motions of the incoherent oscillators outside the Poisson submanifold according to the different set of motion constants, i.e. the non-Poisson initial condition.

In the next step, we have considered two possibilities that make Poisson chimeras attractive or at least remain in a close vicinity of the Poisson submanifold. We have introduced two `perturbations' to the Kuramoto-Sakaguchi phase oscillators on the global two-population network: a nonlocal intra-population topology and an amplitude degree of freedom, i.e. Stuart-Landau planar oscillators. Previously, many authors showed that the OA manifold in the continuum limit is attracting in the long time limit if the system exhibits some type of heterogeneity~\cite{OA2,pikovsky-heteroWS2,OA-attractive}. Considering the WS transformation, it was also shown that a finite-sized system is evolving towards at least a close vicinity of the Poisson submanifold when the system has an suitable heterogeneity, such as nonidentical natural frequencies, noisy oscillators, or experiences a heterogeneous mean-field forcing~\cite{pikovsky-heteroWS3,pikovsky-heteroWS}. We have demonstrated that our two perturbations can be thought of as such a small heterogeneity for the incoherent oscillator population.
Correspondingly, the Lyapunov analysis has revealed that the systems of nonlocally coupled phase oscillators and globally coupled Stuart-Landau amplitude oscillators have (slightly) negative Lyapunov exponents associated with the incoherent population of phase DOFs, and thus an attracting Poisson chimera trajectory~\cite{Laing_SL2010,Laing_SL2019}: Even when starting from non-Poisson ICs, the chimera trajectory evolved towards the Poisson chimera or to a close neighborhood of it in the long time limit.

As a concluding remark, we note that in real world systems heterogeneities of some type will naturally be present so that the Poisson submanifold becomes at least weakly attracting, which underlines the importance of Poisson chimera states.

\section*{Data Availability Statement}
The data that support the findings of this study are available from the corresponding author upon reasonable request.

\begin{acknowledgments}
The authors would like to thank Maximilian Patzauer, Sindre W. Haugland and Felix P. Kemeth for fruitful discussions. This work has been supported by the Deutsche Forschungsgemeinschaft (project KR1189/18 ‘Chimera States and Beyond’)
\end{acknowledgments}

\appendix

\section{\label{append:lyapunov}Lyapunov Analysis}

To study the spectral properties of a chimera trajectory in  state space, we perform a Lyapunov analysis. In this appendix, we review some basic concepts; the detailed descriptions can be found in Refs.~\onlinecite{pikovsky_LE,CLV1,CLV2,CLV3}.

First, our governing equations are represented by a set of autonomous ordinary differential equations. In the general vectorial notation, we consider $\dot{\bold{x}}(t) = \bold{f}(\bold{x}(t))$ with an initial condition $\bold{x}(0)=\bold{x}_0$ where $\bold{x}(t) \in \mathbb{R}^n$ is the dynamical variable, $\bold{f}$ is the vector field, and $n$ is the dimension of the state space. A reference trajectory $\bold{x}_{\textrm{ref}}(t)$ is a solution of the initial value problem, along which we want to study the spectral properties. In our context, it therefore should be a chimera state trajectory. Now we consider the tangent space at each state point along the reference trajectory, wherein the perturbation vector $\delta \bold{x}(t)$ resides, i.e., $\delta \bold{x}(t) \in \mathbf{T}_{\bold{x}_{\textrm{ref}}(t)}(\mathbb{R}^n)$. Those perturbation vectors are governed by the Jacobian matrix of the vector field, evaluated along the reference trajectory, which can be represented as $\delta \dot{\bold{x}}(t) = \bold{J}(t;\bold{x}_{\textrm{ref}}(t) ) \delta \bold{x}(t)$ where the Jacobian matrix is defined by $(\bold{J})_{ij} = \frac{\partial \dot{\bold{x}}_i}{ \partial \bold{x}_j } \big |_{\bold{x}_{\textrm{ref}}(t)} $. From this, we consider the fundamental matrix solution such that $\dot{ \mathcal{O}}(t) = \bold{J}(t;\bold{x}_{\textrm{ref}}(t) ) \mathcal{O}(t)$ with $\mathcal{O}(0)=I_n$; this solution defines $\textit{the tangent linear propagator}$, $\bold{M}(t_0,t) = \mathcal{O}(t)\mathcal{O}^{-1}(t_0)$, of the perturbation vector from a given point in time point to the future time so that $\delta \bold{x}(t) = \bold{M}(t_0,t)\delta \bold{x}(t_0)$~\cite{CLV2}.

Oseledets' theorem~\cite{oseledets, CLV1} tells us that the  limits \eqref{Eq:Oseledets_matrix} exist and share the same real positive eigenvalues denoted by $\mu_1 > \mu_2 > ... > \mu_n$ (Here, we only consider the nondegenerate case). The forward and backward Oseledets matrices are respectively defined by
\begin{flalign}
&\bm{\Xi}^{+}(t) = \lim_{t_2 \rightarrow \infty} \big[ \bold{M}(t,t_2)^{\top}\bold{M}(t,t_2)  \big]^{1/(2(t_2-t))} \notag \\ 
&\bm{\Xi}^{-}(t) = \lim_{t_1 \rightarrow -\infty} \big[ \bold{M}(t_1,t)^{-\top}\bold{M}^{-1}(t_1,t)  \big]^{1/(2(t_1-t))}
\label{Eq:Oseledets_matrix}
\end{flalign}
where $\top$ stands for the transpose of a matrix and $-\top$ for transpose and inverse of it. The forward/backward Oseledets matrix probes the future/past dynamics along the given reference trajectory. Those matrices have the eigenspaces spanned by the so-called \textit{forward/backward Lyapunov vectors} $\bold{d}^{(i)}_{\pm}(t)$. However, these vectors are not covariant under the dynamics, i.e., it does not bear any information on the local expansion/contraction of the perturbation vectors. Nevertheless, we can construct the Oseledets' splitting that decomposes the tangent space according to the local expansion/contraction behavior along the reference trajectory. We define nested subspaces which construct the Oseledets' splitting in following $(\bm{\Gamma}^{(i)}(t))^{+} = \bigoplus_{j=i}^{n}(\bold{U}^{(j)}(t))^{+}$ and $(\bm{\Gamma}^{(i)}(t))^{-} = \bigoplus_{j=1}^{i}(\bold{U}^{(j)}(t))^{-}$ where $(\bold{U}^{(j)}(t))^{\pm}$ are the eigenspaces of the forward/backward Oseledets matrices spanned by $\{\bold{d}_{\pm}^{(j)}(t)\}_{j=1}^{n}$~\cite{CLV1}. Therefore, we have the decomposition of the tangent space such that
$\mathbf{T}_{\bold{x}_{\textrm{ref}}(t)}(\mathbb{R}^n) = \bigoplus_{j=1}^{n}\bm{\Omega}^{(j)}(t)$, where $\bm{\Omega}^{(i)}(t) = (\bm{\Gamma}^{(i)}(t))^{+} \cap (\bm{\Gamma}^{(i)}(t))^{-}$ is called the Oseledets' splitting. This Oseledets' splitting is covariant under the given dynamics in the sense that $\bm{\Omega}^{(i)}(t)=\bold{M}(t_0,t)\bm{\Omega}^{(i)}(t_0)$. The spanning vectors $\{ \bold{v}^{(i)}(t) \}_{i=1}^{n}$ of such Oseledets' splittings are called the $\textit{Covariant Lyapunov Vectors}$ (CLVs)~\cite{kevin} that hold the information on the local expansion/contraction direction of the perturbation vectors since they are norm-independent and also covariant under the dynamics. The exponential rate of such local expansion/contraction along the direction of the CLVs is called \textit{Lyapunov Exponents} (LEs) and defined by 
\begin{equation}
    \Lambda_i = \lim_{t \rightarrow \infty}\frac{1}{t} \textrm{log}\frac{|| \bold{M}(t_0,t) \bold{u}(t_0) ||}{||\bold{u}(t_0)||}
    \label{Eq:LE-definition}
\end{equation} for $\bold{u}(t_0) \in (\bm{\Gamma}^{(i)}(t_0))^{+} \backslash (\bm{\Gamma}^{(i+1)}(t_0))^{+} $ where the nested subspaces are $\mathbb{R}^n = (\bm{\Gamma}^{(1)}(t))^{+} \supset (\bm{\Gamma}^{(2)}(t))^{+} \supset ... \supset (\bm{\Gamma}^{(n)}(t))^{+}$. Hence, the Lyapunov exponents characterize the exponential asymptotic growth rate $|| \bold{M}(t_0,t) \bold{v}^{(i)}(t_0) || \sim ||\bold{v}^{(i)}(t_0)|| \textrm{exp}(\Lambda_i t)$ and the covariant Lyapunov vectors indicate the stable/unstable directions of the perturbation vectors in the state space~\cite{pikovsky_LE}.

\section{\label{append:network-symmetry}Network Symmetry Analysis}

The two-population topology we consider in the main text can, in fact, be seen as a finite-sized network with $2N$ nodes. This holds for both the global and nonlocal intra-population cases. Furthermore, the discrete network symmetries are represented by the automorphism group of a given network~\cite{yscho,pecora1,macarthur,kudose}. Recently, many authors have focused on such network symmetries to investigate the dynamics of various kinds of coupled oscillators on a given finite-sized network with abundant discrete symmetries~\cite{yscho2,pecora2, remote, sjlee1,schaub, krischer_symmetry}. In the following, we exploit the same approach to study the spectral properties of the synchronized population of the chimera states both for the Kuramoto-Sakaguchi phase oscillators and Stuart-Landau amplitude oscillators. In this section, we introduce some important background theories introduced in Refs.~\onlinecite{yscho,pecora1}.

The automorphism group denoted by $\textrm{Aut}(\mathcal{G})$ of a given network $\mathcal{G}$ is a mathematical group consisting of all the automorphisms. An automorphism is a permutation $\sigma$ of the set of nodes that preserve the adjacency relation among the nodes in the way that $A_{ij} = A_{\sigma(i)\sigma(j)}$~\cite{kudose}. Consider the group action under a subgroup $G \leq \textrm{Aut}(\mathcal{G})$. Then, an orbit partition of a given network $\mathcal{G}$ under the subgroup $G$ is a set of orbits defined by $\varphi(G,i) = \{ \sigma(i) | \sigma \in G \}$ which defines a mathematical partition such that $\varphi(G,i) = \varphi(G,j)$ for all $j \in \varphi(G,i)$, and $\varphi(G,i) \cap \varphi(G,j) = \emptyset$ if $j \notin \varphi(G,i)$. This partition of a graph can be a candidate of a cluster synchronization (CS) pattern of a given dynamics on the network~\cite{yscho,yscho2,pecora1,pecora2} since each oscillator in the same orbit should receive the same input from the others.

Let us now consider two different types of governing equations, one of which is called here the Pecora-type equation~\cite{pecora1,pecora2, yscho2} and the other one the Kuramoto-type equation, which describes diffusively coupled oscillators~\cite{yscho, sjlee1, sjlee2}:
\begin{flalign}
&\dot{\bold{x}}_i(t) = \bold{F}(\bold{x}_i(t)) + K \sum_{j=1}^{N}A_{ij}\bold{H}(\bold{x}_j(t)) \notag \\ 
&\dot{\bold{x}}_i(t) = \bold{F}(\bold{x}_i(t)) + K \sum_{j=1}^{N}A_{ij}\bold{H}(\bold{x}_j(t)-\bold{x}_i(t))
\label{Eq:kuramoto-pecora}
\end{flalign}
for $i=1,...,N$ where $\bold{x}_i(t) \in \mathbb{R}^n$ denotes the dynamical variable, $\bold{F(\bold{x})}$ governs the uncoupled dynamics, $\bold{H}(\bold{x})$ the coupling function, and finally $K$ denotes the coupling constant. For a given candidate of CS pattern, we consider the set of all clusters (orbits) $\{ \varphi(i,G) \}_{i=1}^{N} = \{C_m\}_{m=1}^{M}$ where $M$ is the number of clusters, including trivial clusters that have only one oscillator in it. An associated CS dynamics is described by the coarse-grained variables $\{  \bold{s}_m(t) = \bold{x}_i(t) | i \in C_m, 1 \leq m \leq M \}$ under the quotient adjacency matrix $\tilde{A}_{mm'} = \sum_{j\in C_{m'}}A_{ij}$ for an arbitrary node $i \in C_m$, which is nothing but the number of links from an arbitrary node in $C_m$ to all the nodes in $C_{m'}$. Hence, the quotient dynamics of the CS pattern is given by
\begin{flalign}
&\dot{\bold{s}}_m(t) = \bold{F}(\bold{s}_m(t)) + K \sum_{m'=1}^{M}\tilde{A}_{mm'}\bold{H}(\bold{s}_{m'}(t)) \notag \\ 
&\dot{\bold{s}}_m(t) = \bold{F}(\bold{s}_m(t)) + K \sum_{m'=1}^{M}\tilde{A}_{mm'}\bold{H}(\bold{s}_{m'}(t)-\bold{s}_m(t))
\label{Eq:quotient-dynamics}
\end{flalign}
for $m=1,...,M$.

The set of $N$-dimensional orthonormal vectors $\{ \bold{u}_{\kappa}^{(m)} \}_{\kappa=1}^{|C_m|}$ for $m=1,...,M$ called the $\textit{cluster-based coordinates}$ is defined by the following rules~\cite{yscho}: (i) $u_{\kappa i}^{(m)} =0$ if $i \notin C_m$, (ii) for $\kappa=1$, all the nonzero elements of $\bold{u}_{1}^{(m)}$ should be $1/\sqrt{|C_m|}$ that defines the cluster sync-manifold, and (iii) the other vectors $\{\bold{u}_{\kappa}^{(m)}\}_{\kappa=2}^{|C_m|}$ are mutually orthogonal and also to $\bold{u}_{1}^{(m)}$. The cluster-based coordinate transformation can block-diagonalize a relevant matrix such as an adjacency matrix according to the given cluster pattern, which therefore reveals the spectral properties of the dynamics on each cluster~\cite{yscho,yscho2}.

To study the spectral properties of the dynamics on each cluster, for the moment, we only consider the Pecora-type equation in Eqs.~(\ref{Eq:kuramoto-pecora}-\ref{Eq:quotient-dynamics}) and use the given CS pattern as a reference trajectory on which we inflict a small deviation. This, then, yields the coupled variational equations for all the clusters
\begin{equation}
    \delta \dot{\bold{x}}_i(t) = D\bold{F}(\bold{s}_m)\delta \bold{x}_i +K \sum_{m'=1}^{M}\sum_{j \in C_{m'}}A_{ij}D\bold{H}(\bold{s}_{m'})\delta \bold{x}_j
    \label{Eq:coupled-variational}
\end{equation}
for $i=1,...,N$ where $\delta \bold{x}_i(t)=\bold{x}_i(t)-\bold{s}_m(t)$ for $i \in C_m$ and $D\bold{F}$ and $D\bold{H}$ indicate the Jacobian matrices of the given dynamical functions. Notice that each variational equation in Eq.~(\ref{Eq:coupled-variational}) is coupled to all the others through the given adjacency matrix. However, if we see this in the cluster-based coordinates by following $\bm{\eta}^{(m)}_{\kappa}  = \sum_{i \in C_m}u_{\kappa i}^{(m)} \delta \bold{x}_i$ for $m=1,...,M $ and $\kappa=2,...,|C_m|$ where $\bm{\eta}_{\kappa}^{(m)}$ for $\kappa \geq 2$ represents the perturbation of the transverse direction to the cluster $C_m$, we get the variational equation for that cluster, independent of the other clusters, provided that the given cluster $C_m$ is non-intertwined with the others (see Supplemental Material in Ref.~\onlinecite{yscho}). Therefore, the $|C_m|-1$ transversal variational equations of the cluster $C_m$ both for the Pecora-type and Kuramoto-type are given by~\cite{yscho}
\begin{flalign}
\dot{\bm{\eta}}_{\kappa}^{(m)}(t) =&  \bigg[ D\bold{F}(\bold{s}_m)+K\lambda_{\kappa}^{(m)}D\bold{H}(\bold{s}_m) \bigg] \bm{\eta}_{\kappa}^{(m)}(t)  \notag \\ 
\dot{\bm{\eta}}_{\kappa}^{(m)}(t) =&  \bigg[ D\bold{F}(\bold{s}_m)-K\sum_{m'=1}^{M}\tilde{A}_{mm'}D\bold{H}(\bold{s}_{m'}-\bold{s}_m) \notag \\
&+K\lambda_{\kappa}^{(m)}D\bold{H}(0) \bigg] \bm{\eta}_{\kappa}^{(m)}(t)
\label{Eq:transverse-variational}
\end{flalign} for $\kappa=2,...,|C_m|$, where $\lambda_{\kappa}^{(m)}$ is the eigenvalue of the adjacency matrix corresponding to the cluster $C_m$ with the eigenvector $\bold{u}_{\kappa}^{(m)}$ of the adjacency matrix. From those transversal variational equations, we can investigate the spectral information on the transverse direction of each cluster along our chimera states.

\section{\label{sec:mathematical-LE} Lyapunov Exponents based on Network Symmetry-induced Cluster Patterns}

\subsection{\label{subsec:KS-global-matheLE} Kuramoto-Sakaguchi Phase Oscillators}

As a first step, we identify the cluster-synchronization (CS) pattern corresponding to the chimera state on the two-population network by assigning one of the two populations to the synchronized oscillators, and the other one to the incoherent oscillators. 
The population of the $N$ perfectly synchronized oscillators can be thought of as just one giant cluster, which we denote by $C_0$,  whereas each incoherent oscillator in the other population is treated as a trivial cluster denoted by $C_m$ with $m=1,...,N$. This gives us the corresponding cluster-based coordinates $U^\top=[ \bold{u}_{1}^{(0)}, \bold{u}_{1}^{(1)},...,\bold{u}_{1}^{(N)}, \bold{u}_{2}^{(0)},...,\bold{u}_{N}^{(0)} ]$ for the chimera pattern~\cite{yscho}. Here, $\bold{u}_{1}^{(0)}$ indicates the direction along the synchronized cluster $C_0$ of the chimera state, so that $u_{1 j}^{(0)}=\frac{1}{\sqrt{N}}$ for $j \in C_0$ and $u_{1 j}^{(0)}=0$ if $j \notin C_0$. For the transverse directions, we obtain $\sum_{j \in C_0}u_{\kappa j}^{(0)}=0$ and $u_{\kappa j}^{(0)}=0$ if $j \notin C_0$ for $\kappa=2,...,N$. Finally, for the incoherent trivial clusters we have $u_{1 j}^{(m)} = 1$ if $j \in C_m$ and $u_{1 j}^{(m)}=0$ otherwise, with $m=1,...,N$. Note that all the cluster-based coordinate vectors should be mutually orthonormalized. An example of a possible candidate of the cluster-based coordinates is~\cite{github}
\begin{equation}
U^\top=
\begin{pmatrix}
    
    \frac{1}{\sqrt{N}} & \vline &  &  &  & \vline & & &  \\
    \vdots & \vline &  & \mathrm{O}_{N, N} &  & \vline & &\mathrm{P}&  \\ 
    \frac{1}{\sqrt{N}} & \vline &  & &  & \vline & & &  \\ \cline{1-9}
   
   0 & \vline &  & &  & \vline & & &  \\
   \vdots & \vline &  & \mathrm{D}&  & \vline & &\mathrm{O}_{N,N-1} & \\
   0 & \vline &  & &  & \vline & & &  \\
    
    \end{pmatrix} 
    \label{Eq:cluster-based}
\end{equation} where the first column $\bold{u}^{(0)}_1 = [\frac{1}{\sqrt{N}},\cdots,\frac{1}{\sqrt{N}},0,\cdots,0]^\top$ indicates the sync-manifold direction, $\mathrm{D} = \textrm{diag}(1,...,1) \in \mathbb{R}^{N\times N}$ indicating the incoherent trivial clusters, each $\mathrm{O}$ is a zero-matrix, and $\mathrm{P}  \in \mathbb{R}^{N\times N-1}$ representing the directions transverse to $C_0$, can be chosen to satisfy orthonormality and transversality such as
\begin{equation}
\mathrm{P}=
\begin{pmatrix}
    \frac{N-1}{\sqrt{N(N-1)}} & 0 & 0 & 0 \\
    -\frac{1}{\sqrt{N(N-1)}} & \frac{N-2}{\sqrt{(N-1)(N-2)}} &0 &0 \\
   -\frac{1}{\sqrt{N(N-1)}} & -\frac{1}{\sqrt{(N-1)(N-2)}} & \ddots & \vdots \\
    \vdots & \vdots & \ddots & \frac{1}{\sqrt{2\cdot 1}} \\
    -\frac{1}{\sqrt{N(N-1)}} & -\frac{1}{\sqrt{(N-1)(N-2)}} & \cdots & -\frac{1}{\sqrt{2\cdot 1}}
    
    \end{pmatrix} 
    \notag
\end{equation} The cluster-based coordinates decouple the variational equations according to the given CS pattern, as demonstrated in Appendix.~\ref{append:network-symmetry} and in the Supplemental Material of Ref.~\onlinecite{yscho}. Our case is rather simple since our chimera state has only one nontrivial cluster for the synchronized oscillators.

Considering the two-population topology as one large network consisting of $2N$ nodes with appropriately defined coupling weights, and describing a chimera state by a CS pattern defined above $\{ C_m \}_{m=0}^{N}$, the Lyapunov exponents corresponding to the synchronized cluster $C_0$ can be analytically estimated. According to this approach, the governing equation can be written as
\begin{equation}
\frac{d}{dt} \phi_i(t) = \mathrm{F}(\phi_i(t))+\sum_{j=1}^{2N}K_{ij}B^{(c)}_{ij}\mathrm{H}(\phi_j(t)-\phi_i(t))
\label{Eq:KS-global-ansatz}
\end{equation}
for $i=1,...,2N$ where the uncoupled dynamics is $\mathrm{F}(\phi)=-\frac{\mu}{N}\textrm{sin}\alpha$ (here, just a constant) and the coupling function is $\mathrm{H}(x)=\textrm{sin}(x-\alpha)$. This is nothing but the Kuramoto-type equation discussed in Eq.~(\ref{Eq:kuramoto-pecora}). The adjacency matrix $B^{(c)}_{ij} \in \mathbb{R}^{2N \times 2N}$ stands for the complete graph with $2N$ nodes, and the coupling weights are defined by $K_{ij} = \frac{\mu}{N}$ if $i,j$ belong to the same population, and $K_{ij}=\frac{\nu}{N}$ if $i,j$ belong to different populations, respectively, for $i,j=1,...,2N$. From the CS pattern $\{ C_m \}_{m=0}^{N}$, the quotient adjacency matrix is given as
\begin{equation}
\tilde{A} =
\begin{pmatrix}
    N-1 & \vline &  1  & \cdots &  1     \\ \cline{1-5}
    N & \vline &   &      &    \\
    \vdots & \vline &  & A^{(c)} & \\
    N & \vline &  &   & 
    \end{pmatrix}
    \label{Eq:global-quotient-adjacency-matrix}
\end{equation} where $A^{(c)}  \in \mathbb{R}^{N \times N}$ is the adjacency matrix of the complete graph with $N$ nodes that describes the global intra-population coupling. Note that the quotient adjacency matrix in Eq.~(\ref{Eq:global-quotient-adjacency-matrix}) is an $\mathbb{R}^{(N+1) \times (N+1)}$ matrix and the index is taken from $0$ to $N$ for the sake of simplicity: $\tilde{A}_{mm'}$ for $m,m'=0,1,...,N$. Therefore, we obtain the (coarse-grained) quotient dynamics corresponding to our chimera pattern from Eq.~(\ref{Eq:quotient-dynamics}) with the CS variables denoted by $s_0(t)=\phi_i(t)$ (sync., $C_0$) and $s_m(t)=\phi_{i+N}(t)$ (incoh., $C_m$) for $m=i=1,...,N$:
\begin{flalign}
\dot{s}_0(t) &= \mathrm{F}(s_0(t)) +\frac{\mu}{N} \mathrm{H}(0)\tilde{A}_{00} +\frac{\nu}{N}\sum_{m'=1}^{N}\tilde{A}_{0m'}\mathrm{H}(s_{m'}(t)-s_0(t)) \notag \\ &= -\mu\textrm{sin}\alpha +\frac{\nu}{N}\sum_{m'=1}^{N}\textrm{sin}(s_{m'}(t)-s_0(t)-\alpha)
\label{Eq:KS-sync-cluster}
\end{flalign} for the synchronized cluster ($C_0$) where the quotient adjacency matrix $\tilde{A}_{00}=N-1$ and $\tilde{A}_{0m'}=1$ for $m'=1,...,N$, and $\mathrm{H}(0)=-\textrm{sin}\alpha$. The quotient governing equations of the $N$ trivial clusters ($C_1,...,C_N$) for the incoherent population read
\begin{flalign}
\dot{s}_m(t)&= \mathrm{F}(s_m) + \frac{\nu}{N}\tilde{A}_{m0} \mathrm{H}(s_0-s_m) +\frac{\mu}{N}\sum_{m'=1}^{N}\tilde{A}_{mm'}\mathrm{H}(s_{m'}-s_m)  \notag \\ &= \nu \textrm{sin}(s_0-s_m-\alpha) +\frac{\mu}{N}\sum_{m'=1}^{N} \textrm{sin}(s_{m'}-s_m-\alpha)
\label{Eq:KS-incoherent-clusters}
\end{flalign} for $m=1,...,N$. From the quotient dynamics, we consider the variational equations of the synchronized oscillators around the CS pattern as
\begin{flalign}
\delta\dot{\phi}_i &= D\mathrm{F}(s_0)\delta\phi_i - \sum_{m'=0}^{N}\sum_{j \in C_{m'}}K_{ij}B^{(c)}_{ij}D\mathrm{H}(s_{m'}-s_0)\delta \phi_i \notag \\ &+ \sum_{m'=0}^{N}\sum_{k \in C_{m'}}K_{ik}B^{(c)}_{ik}D\mathrm{H}(s_{m'}-s_0)\delta \phi_k \notag \\
&= D\mathrm{F}(s_0)\delta\phi_i - \frac{\mu}{N}\tilde{A}_{00}D\mathrm{H}(0)\delta\phi_i \notag \\ &-\frac{\nu}{N}\sum_{m'=1}^{N}\tilde{A}_{0m'}D\mathrm{H}(s_{m'}-s_0)\delta\phi_i +\frac{\mu}{N}\sum_{k\in C_0}B^{(c)}_{ik}D\mathrm{H}(0)\delta\phi_k \notag\\ &+\frac{\nu}{N}\sum_{m'=1}^{N}\sum_{k\in C_{m'}}B^{(c)}_{ik}D\mathrm{H}(s_{m'}-s_0)\delta\phi_k
\label{Eq:variation}
\end{flalign} for each $i \in C_0$ where the deviation around the CS pattern is $\delta \phi_i(t) = \phi_i(t)-s_m(t)$ for $i\in C_m$ and $m=0,1,...,N$. Next, we want to obtain Eq.~(\ref{Eq:variation}) in the cluster-based coordinate defined in Eq.~(\ref{Eq:cluster-based}). The transverse variations can be written as $\eta_{\kappa}^{(0)}(t) = \sum_{i \in C_0}u_{\kappa i}^{(0)} \delta \phi_i(t)$ with $U=[ \bold{u}_{1}^{(0)}, \bold{u}_{1}^{(1)},...,\bold{u}_{1}^{(N)}, \bold{u}_{2}^{(0)},...,\bold{u}_{N}^{(0)} ]^\top$. Then, the variational equations transverse to the sync-cluster $C_0$ read
\begin{flalign}
\dot{\eta}_{\kappa}^{(0)} &= \sum_{i \in C_0}u_{\kappa i}^{(0)} \delta \dot{\phi}_i(t) = \sum_{i \in C_0} u_{\kappa i}^{(0)} \Bigg( D\mathrm{F}(s_0) \notag \\ 
& - \frac{\mu}{N}\tilde{A}_{00}D\mathrm{H}(0)-\frac{\nu}{N}\sum_{m'=1}^{N}\tilde{A}_{0m'}D\mathrm{H}(s_{m'}-s_0) \Bigg)\delta \phi_i \notag \\ & +\frac{\mu}{N} D\mathrm{H}(0)\sum_{k \in C_0}\sum_{i \in C_0}  u_{\kappa i}^{(0)} B_{ik}^{(c)}\delta \phi_{k} \notag \\ &+ \frac{\nu}{N}\sum_{m'=1}^{N}\sum_{i\in C_0}\sum_{k\in C_{m'}}u_{\kappa i}^{(0)} B^{(c)}_{ik}D\mathrm{H}(s_{m'}-s_0)\delta\phi_k \notag \\
&=\Bigg( D\mathrm{F}(s_0)-\frac{\mu}{N}\tilde{A}_{00}D\mathrm{H}(0)-\frac{\nu}{N}\sum_{m'=1}^{N}\tilde{A}_{0m'}D\mathrm{H}(s_{m'}-s_0) \Bigg) \eta_{\kappa}^{(0)} \notag \\
&+\frac{\mu}{N}D\mathrm{H}(0)\sum_{k \in C_0}\sum_{i \in C_0}\sum_{\kappa'=1}^{|C_{0}|}  u_{\kappa i}^{(0)} B_{ik}^{(c)} u_{\kappa' k}^{(0)} \eta_{\kappa'}^{(0)} \notag \\ & + \frac{\nu}{N}\sum_{m'=1}^{N}\sum_{i\in C_0}\sum_{k\in C_{m'}}\sum_{\kappa'=1}^{|C_{m'}|}u_{\kappa i}^{(0)} B^{(c)}_{ik}u_{\kappa' k}^{(m')}D\mathrm{H}(s_{m'}-s_0)\eta_{\kappa'}^{(m')}
\label{Eq:trans-variation}
\end{flalign} for $\kappa=2,...,N$. As shown in Ref.~\onlinecite{yscho}, the cluster-based coordinates can block-diagonalize the adjacency matrix $B^{(c)}$ according to the CS pattern so that the block corresponding to the sync-cluster $C_0$ can be represented by the matrix $\textrm{diag}(\lambda_{2}^{(0)},\lambda_{3}^{(0)},...,\lambda_{N}^{(0)}) \in \mathbb{R}^{(N-1)\times(N-1)}$ and the off-diagonal blocks are zero. This, in turn, means that the last term in Eq.~(\ref{Eq:trans-variation}) should be zero and $\sum_{i\in C_0}\sum_{k \in C_0} u_{\kappa i}^{(0)} B_{ik}^{(c)} u_{\kappa' k}^{(0)} = \lambda_{\kappa}^{(0)}\delta_{\kappa \kappa'}$ for $\kappa, \kappa'=2,...,N$ where $\lambda_{\kappa}^{(0)}$ are the eigenvalues of the adjacency matrix since $\bold{u}^{(0)}_{\kappa}$ for $\kappa=2,...,N$ can be chosen to be the eigenvectors of the adjacency matrix~\cite{yscho,github}. Hence, the variational equations transverse to the sync-manifold are given by
\begin{flalign}
\dot{\eta}_{\kappa}^{(0)} &=   \Bigg[ D\mathrm{F}(s_0) - \frac{\mu}{N}\tilde{A}_{00}D\mathrm{H}(0)  +\frac{\mu}{N} \lambda_{\kappa}^{(0)} D\mathrm{H}(0)  \notag \\ & - \frac{\nu}{N}\sum_{m'=1}^{N}\tilde{A}_{0m'}D\mathrm{H}(s_{m'}-s_0) \Bigg] \eta_{\kappa}^{(0)} \notag \\&= \Bigg[ -\frac{\mu}{N}(N-1) \textrm{cos}\alpha +\frac{\mu}{N} \lambda_{\kappa}^{(0)}  \textrm{cos}\alpha  \notag \\
&- \frac{\nu}{N}\sum_{m'=1}^{N} \textrm{cos}(s_{m'}-s_0-\alpha)  \Bigg] \eta_{\kappa}^{(0)}
\label{Eq:transverse-phaseonly}
\end{flalign} for $\kappa=2,...,N$. Notice that for the global intra- and inter- population network, the eigenvalues in Eq.~(\ref{Eq:transverse-phaseonly}) $\lambda_{\kappa}^{(0)}=-1$ for all $\kappa=2,...,N$. Consider as an example the system with $N=4$. Its block-diagonalized adjacency matrix reads~\cite{github}
\begin{equation}
UB^{(c)}U^{-1}=
\begin{pmatrix}
   3 & 2 & 2 & 2 & 2 & \vline&  &  &  \\
   2 & 0 & 1 & 1 & 1 & \vline& &  &  \\
  2 & 1 & 0 & 1 & 1 & \vline& & \mathrm{O}_{5,3} &  \\
  2 & 1 & 1 & 0 & 1& \vline& &  &  \\   
   2 & 1 & 1 & 1 & 0 & \vline& &  &  \\ \cline{1-9}
    &  &  &  &  & \vline&-1 & 0 & 0 \\
    &  & \mathrm{O}_{3,5} &  &  & \vline&0 & -1 & 0 \\   
    &  &  &  &  &\vline& 0 & 0 & -1\\   
    \end{pmatrix} 
    \notag
\end{equation} where the lower-right block corresponds to the sync-cluster $C_0$ and we obtain $\lambda_{\kappa}^{(0)}=-1$ for all $\kappa$ for our global intra- and inter- population topology. Hence, if we consider the summation term in Eq.~(\ref{Eq:transverse-phaseonly}) as an external forcing field~\cite{abrams_chimera2016}, then it gives approximated values of the $(N-1)$-fold degenerate transverse LEs in Eq.~(\ref{Eq:trans-LE-phase-global}).

To estimate the Lyapunov exponent along the sync-manifold for the synchronized population $\Lambda_{\textrm{perturb}}^{(0)}$, the perturbation should be performed along the sync-manifold. This means we obtain the variational equation when the small perturbation $s_0(t) \rightarrow s_0(t) + \delta s_0(t)$ where $|\delta s_0 (t)| \ll 1$ is applied to  Eq.~(\ref{Eq:KS-sync-cluster}).
\begin{flalign}
\frac{d}{dt} \delta s_0(t) &= D\mathrm{F}(s_0) \delta s_0(t) + \frac{\nu}{N}\sum_{j=1}^{N}D\mathrm{H}(s_j-s_0)(-\delta s_0) \notag\\
&= - \Bigg[ \frac{\nu}{N}\sum_{m'=1}^{N}\textrm{cos}(s_{m'}-s_0-\alpha) \Bigg] \delta s_0(t) \label{Eq:KS-global-sync-along-varEQ}
\end{flalign} Then, we obtain Eq.~(\ref{Eq:KS-sync-perturb-LE}) provided that $Z=\frac{\nu}{N}\sum_{m'=1}^{N}\textrm{cos}(s_{m'}-s_0-\alpha)$ is regarded as external forcing function.

For the Kuramoto-Sakaguchi phase oscillators in the nonlocal intra-population network, we use the same ansatz introduced above where we treated the chimera state as a CS pattern dynamics. We again start the analysis with the governing equation that, however, now contain the nonlocal adjacency matrix
\begin{flalign}
\frac{d }{dt}\phi_i(t) = \mathrm{F}(\phi_i(t))+\sum_{j=1}^{2N}K_{ij}B^{(\textrm{n})}_{ij}\mathrm{H}(\phi_j(t)-\phi_i(t))
\label{Eq:nonlocal-phase-ansatz}
\end{flalign} for $i=1,...,2N$ where $\mathrm{F}(\phi)$, $K_{ij}$, and $\mathrm{H}(x)$ are the same as defined in Eq.~(\ref{Eq:KS-global-ansatz}). The matrix $B^{(\textrm{n})} \in \mathbb{R}^{2N \times 2N}$, which defines the global inter- and nonlocal intra-population network, is given by
\begin{equation}
B^{\textrm{(n)}}=
\begin{pmatrix}
    A &\vline & J_N   \\ \cline{1-3}
    J_N&\vline & A
    \end{pmatrix} \in \mathbb{R}^{2N\times 2N} \notag
\end{equation} where $A \in \mathbb{R}^{N \times N}$ is defined in Eq.~(\ref{Eq:nonlocal-adjacency-matrix}) and $J_N \in \mathbb{R}^{N \times N}$ is the unit matrix whose elements are all $1$. In this ansatz, the quotient adjacency matrix is given by
\begin{equation}
\tilde{A}^{\textrm{(n)}}=
\begin{pmatrix}
    N-2 &\vline & & 1 ~~ \dots ~~ 1  &   \\ \cline{1-4}
    N&\vline &   &      &    \\
    \vdots&\vline &  & A & \\
    N&\vline &       &   & 
    \end{pmatrix} 
    \label{Eq:quotient-adjacency-matrix-nonlocal}
\end{equation} wherein the terms $\tilde{A}^{\textrm{{(n)}}}_{00}=N-2$ and $\tilde{A}^{\textrm{{(n})}}_{ij}=A_{mm'}$ for $m,m'=1,...,N$ ensuring that the intra-population topology is not global but nonlocal. From $\tilde{A}^{\textrm{(n)}}$, we obtain the quotient dynamics according to the CS pattern describing our chimeras with the variables $s_0(t)=\phi_i(t)$ (sync.) and $s_m(t)=\phi_{i+N}(t)$ (incoh.) for $i=m=1,...,N$:
\begin{flalign}
\frac{d s_0}{dt} &= \mathrm{F}(s_0) + \frac{\mu}{N}\tilde{A}^{\textrm{(n)}}_{00}\mathrm{H}(0) + \frac{\nu}{N}\sum_{m'=1}^{N}\tilde{A}^{\textrm{(n)}}_{0m'} \mathrm{H}(s_{m'}-s_0) \notag \\ &= -\frac{\mu}{N}(N-1)\textrm{sin}\alpha +\frac{\nu}{N}\sum_{m'=1}^{N}\textrm{sin}(s_{m'}-s_0-\alpha)
\label{Eq:quotient_nonlocal-phase-sync}
\end{flalign} for the synchronized population, and
\begin{flalign}
\frac{d s_m}{dt} &= \mathrm{F}(s_m) +\frac{\nu}{N}\tilde{A}^{\textrm{(n)}}_{m0}\mathrm{H}(s_0-s_m) +\frac{\mu}{N}\sum_{m'=1}^{N}\tilde{A}^{(n)}_{mm'} \mathrm{H}(s_{m'}-s_m) \notag\\&= -\frac{\mu}{N}\textrm{sin}\alpha +\nu\textrm{sin}(s_0-s_m-\alpha) \notag \\ &+\frac{\mu}{N}\sum_{m'=1}^{N}A_{mm'}\textrm{sin}(s_{m'}-s_m-\alpha) \notag \\
&= \tilde{\omega}_m(t) + \nu\textrm{sin}(s_0-s_m-\alpha) + \frac{\mu}{N}\sum_{m'=1}^{N}\textrm{sin}(s_{m'}-s_m-\alpha)
\label{Eq:quotient_nonlocal-phase-incoherent}
\end{flalign} where $\tilde{\omega}_m(t)=-\frac{\mu}{N}\textrm{sin}(s_{m+N/2}-s_m-\alpha)$ with $\tilde{A}^{\textrm{(n)}}_{0m}=1$ and $\tilde{A}^{\textrm{(n)}}_{m0}=N$ for $m=1,...,N$ for the incoherent trivial clusters.

As seen in Sec.~\ref{subsec:KS-nonlocal-LEfull}, there are $N-1$ transverse Lyapunov exponents consisting of two different values. This splitting of the values of $\Lambda_{\textrm{trans}}^{(0)}$ is due to the two different eigenvalues of the nonlocal adjacency matrix. As clear from Eq.~(\ref{Eq:transverse-variational}), one has to consider the eigenvalues of the adjacency matrix associated with the cluster-based vector, which are the eigenvectors of the adjacency matrix $\bold{u}_{\kappa}^{(m)}$, to obtain the transverse variational equations. For the global topology discussed above, these eigenvalues $\lambda_{\kappa}^{(0)} =-1$ are the same for $\kappa=2,...,N$. In contrast, the nonlocal adjacency matrix has two different eigenvalues: $\lambda_{\kappa}^{(0)} =0$ for $\kappa= 2,...,N/2+1$ and $\lambda_{\kappa}^{(0)}=-2$ for $\kappa=N/2+2,...,N$~\cite{github}. This leads to two different variational equations with the same method in Eqs.~(\ref{Eq:trans-variation}-\ref{Eq:transverse-phaseonly})
\begin{flalign}
\dot{\eta}_{\kappa}^{(0)} &= \Bigg[ D\mathrm{F}(s_0) - \frac{\mu}{N}\tilde{A}^{(n)}_{00}D\mathrm{H}(0)+\frac{\mu}{N}\lambda_{\kappa}^{(0)} D\mathrm{H}(0) \notag \\ &-\frac{\nu}{N}\sum_{m'=1}^{N}\tilde{A}^{(n)}_{0m'}D\mathrm{H}(s_{m'}-s_0) \Bigg] \eta_{\kappa}^{(0)} \notag \\ &= \Bigg[ -\frac{\mu}{N} (N-2) \textrm{cos}\alpha +\frac{\mu}{N}\lambda_{\kappa}^{(0)}\textrm{cos}\alpha - \frac{\nu}{N}\mathrm{Z}
 \Bigg] \eta_{\kappa}^{(0)}
\label{Eq:transverse-nonlocal-phase}
\end{flalign} for $\kappa =2,...,N$. Therefore,
Eq.~(\ref{Eq:transverse-nonlocal-phase}) yields two different groups of degenerate Lyapunov exponents transverse to the sync-manifold 
\begin{flalign}
   \Lambda_{\textrm{trans},\kappa}^{(0)} &= -\frac{\mu}{N} (N-2) \textrm{cos}\alpha +\frac{\mu}{N}\lambda_{\kappa}^{(0)}\textrm{cos}\alpha - \frac{\nu}{N}\mathrm{Z} \notag \\
   &= \begin{dcases}
    -\frac{\mu}{N}(N-2)\textrm{cos}\alpha -\frac{\nu}{N}\mathrm{Z} <0, & \kappa =  2,...,N/2+1 \\ \\
     - \mu\textrm{cos}\alpha -\frac{\nu}{N}\mathrm{Z}<0, & \kappa=N/2+2,...,N
  \end{dcases}
  \label{Eq:KS-nonlocal-trans-LE}
\end{flalign} provided that $\mathrm{Z}$ is treated as an external forcing field. Also, there is another LE of the synchronized population, which arises from a perturbation along the sync-manifold. Here, a small perturbation $s_0 \rightarrow s_0 + \delta s_0$ is imposed on Eq.~(\ref{Eq:quotient_nonlocal-phase-sync}) where $|\delta s_0| \ll 1$. This perturbation gives $\Lambda_{\textrm{sync}}^{(0)} = -\frac{\nu}{N}\mathrm{Z}<0$ strongly depending on the motion of the incoherent oscillators.

\subsection{\label{subsec:KS-global-matheLE} Stuart-Landau Planar Oscillators}

Let us consider the spectra corresponding to the amplitude DOFs in more detail. Using the corresponding ansatz as above, the evolution of the amplitude DOFs can be expressed as
\begin{flalign}
\frac{d r_i(t)}{dt} = \mathrm{F}^{\textrm{(amp)}}(r_i(t)) +\sum_{j=1}^{2N} K_{ij}^{\textrm{(amp)}}B^{(c)}_{ij} \mathrm{H}^{\textrm{(amp)}}(r_j(t))
\label{amplitudeDOF-ansatz}
\end{flalign}
for $i=1,...,2N$, where $\mathrm{F}^{\textrm{(amp)}}(r) = \epsilon^{-1}(1-r^2)r + \frac{\mu}{N}r\textrm{cos}\alpha$ and $\mathrm{H}^{\textrm{(amp)}}(r)=r$. Here, we regard the phase variables as external forcing functions, which means we define the coupling weight in Eq.~(\ref{amplitudeDOF-ansatz}) as $K_{ij}^{\textrm{(amp)}}=\frac{\mu}{N}\textrm{cos}(\phi_j-\phi_i-\alpha)$ if $i,j$ belong to the same population and $K_{ij}^{\textrm{(amp)}}=\frac{\nu}{N}\textrm{cos}(\phi_j-\phi_i-\alpha)$ if $i,j$ belong to the different populations. This equation is a Pecora-type equation (cf. Eq.~(\ref{Eq:kuramoto-pecora})), and the amplitude Lyapunov exponents can be approximated as follows.

According to the chimera CS pattern dynamics introduced in Sec.~\ref{subsec:KS-global-matheLE}, we denote the amplitude degrees of freedom by $r_i(t) = R_0(t)=1$ for the synchronized population and $r_{i+N}(t)=R_m(t)$ for the incoherent one, and, correspondingly, the phase DOFs by $s_0(t)=\phi_i(t)$ (sync.) and $s_m(t)=\phi_{i+N}(t)$ (incoh.) for $i=m=1,...,N$. Then, the quotient dynamics of the amplitude DOFs for the synchronized population with the quotient adjacency matrix in Eq.~(\ref{Eq:global-quotient-adjacency-matrix}) is governed by

\begin{flalign}
\frac{dR_0}{dt}&= \mathrm{F}^{(\textrm{amp})}(R_0)+\frac{\mu}{N}\tilde{A}_{00}\mathrm{H}^{(\textrm{amp})}(R_0)\textrm{cos}\alpha \notag\\&+\frac{\nu}{N}\sum_{m'=1}^{N}\tilde{A}_{0m'}\mathrm{H}^{(\textrm{amp})}(R_{m'})\textrm{cos}(s_{m'}-s_0-\alpha) \notag \\ &= \bigg( \epsilon^{-1}(1-R^2_0) +\frac{\mu}{N}\textrm{cos}\alpha \bigg)R_0 +\frac{\mu}{N}(N-1)R_0 \textrm{cos}\alpha \notag \\
&+\frac{\nu}{N}\sum_{m'=1}^{N}R_{m'} \textrm{cos}(s_{m'}-s_0-\alpha).
\label{amplitudeDOF-quotient}
\end{flalign} Considering a small deviation around the CS dynamics, i.e., $\delta r_i (t) = r_i(t) - R_m(t) $ for $i \in C_m$ for $m=0,1,...,N$, we obtain the coupled variational equations as
\begin{flalign}
\delta\dot{r}_i (t) &= D\mathrm{F}^{(\textrm{amp})}(R_0)\delta r_{i} + \frac{\mu}{N}\mathrm{C}_{00}\sum_{k\in C_0}B^{(c)}_{ik}D\mathrm{H}^{(\textrm{amp})}(R_0)\delta r_{k} \notag \\ &+ \frac{\nu}{N}\sum_{m'=1}^{N}\sum_{k\in C_{m'}}B^{(c)}_{ik}D\mathrm{H}^{(\textrm{amp})}(R_{m'})\mathrm{C}_{m'0}\delta r_k \notag
\end{flalign} for each $i \in C_0$ and $\mathrm{C}_{m'm}=\textrm{cos}(s_{m'}-s_m-\alpha)$ for $m,m'=0,...,N$. Then, viewing these in the cluster-based coordinates with $\xi^{(0)}_{\kappa}(t) = \sum_{i \in C_0} u_{\kappa i}^{(0)}\delta r_i(t)$ for $\kappa=2,...,N$, the transversal variational equations in Eq.~(\ref{Eq:transverse-variational}) read 
\begin{flalign}
\dot{\xi}_{\kappa}^{(0)} &= \sum_{i \in C_0} u_{\kappa i}^{(0)}\delta \dot{r}_i(t)= D\mathrm{F}^{(\textrm{amp})}(R_0)\sum_{i \in C_0} u_{\kappa i}^{(0)} \delta r_{i} \notag \\
& + \frac{\mu}{N} D\mathrm{H}^{(\textrm{amp})}(R_0)\mathrm{C}_{00} \sum_{i \in C_0}\sum_{k \in C_0}u_{\kappa i}^{(0)} B^{(c)}_{ik} \delta r_k \notag\\&+ \frac{\nu}{N}\sum_{i \in C_0}\sum_{m'=1}^{N}\sum_{k\in C_{m'}}u_{\kappa i}^{(0)}B^{(c)}_{ik}D\mathrm{H}^{(\textrm{amp})}(R_{m'}) \mathrm{C}_{m'0} \delta r_k \notag\\
&=  D\mathrm{F}^{(\textrm{amp})}(R_0) \xi_{\kappa}^{(0)} \notag \\
&+\frac{\mu}{N}D\mathrm{H}^{(\textrm{amp})}(R_0)\mathrm{C}_{00} \sum_{i \in C_0}\sum_{k \in C_0}\sum_{\kappa'=1}^{|C_0|}u_{\kappa i}^{(0)} B^{(c)}_{ik} u_{\kappa' k}^{(0)} \xi_{\kappa'}^{(0)} \notag \\ &+ \frac{\nu}{N}\sum_{m'=1}^{N}\sum_{i \in C_0}\sum_{k\in C_{m'}}\sum_{\kappa' = 1}^{|C_{m'}|}u_{\kappa i}^{(0)}B^{(c)}_{ik}u_{\kappa' k}^{(m')} \xi_{\kappa'}^{(m')} \mathrm{C}_{m'0}D\mathrm{H}^{(\textrm{amp})}(R_{m'})
\notag
\end{flalign} where the last term is zero and since the adjacency matrix is block-diagonalizd in the cluster-based coordinates $\sum_{i\in C_0}\sum_{k \in C_0} u_{\kappa i}^{(0)} B_{ik}^{(c)} u_{\kappa' k}^{(0)} = \lambda_{\kappa}^{(0)}\delta_{\kappa \kappa'}$ for $\kappa=2,...,N$. Hence, the $N-1$  variational equations transversal to the sync-manifold are given by \begin{flalign}
\dot{\xi}_{\kappa}^{(0)} &= \Bigg[ D\mathrm{F}^{(\textrm{amp})}(R_0)  + \frac{\mu}{N}\textrm{cos}\alpha \lambda_{\kappa}^{(0)}D\mathrm{H}^{(\textrm{amp})}(R_0) \Bigg]\xi_{\kappa}^{(0)} \notag\\&= \bigg[ \epsilon^{-1}(1-3R^2_0)+\frac{\mu}{N}(1+\lambda_{\kappa}^{(0)})\textrm{cos}\alpha \bigg] \xi_{\kappa}^{(0)}
\label{Eq:trans-varial-amplitude}
\end{flalign} Here, the $\lambda_{\kappa}^{(0)}=-1$ since they are the same as for the global intra-population network (Eq.~(\ref{Eq:transverse-phaseonly})). Thus, with Eq.~(\ref{Eq:trans-varial-amplitude}) we obtain the approximate values of the $(N-1)$-fold degenerate transverse Lyapunov exponents in the amplitude DOFs as $\Lambda_{\textrm{trans},\kappa}^{(\textrm{amp},0)} \approx \epsilon^{-1}(1-3R^2_0)<0 $ in Eq.~(\ref{Eq:SL-global-Amp-LE-trans})
for $\kappa=2,...,N$. 

Next, to estimate the Lyapunov exponent associated with the perturbation along the sync-manifold in the amplitude DOFs, we perform a small perturbation along the sync-manifold  $R_0(t) \rightarrow R_0(t)+\delta R_0(t)$ with $|\delta R_0| \ll 1$ in Eq.~(\ref{amplitudeDOF-quotient}) and obtain
\begin{flalign}
\delta \dot{R}_0 (t) &= D\mathrm{F}^{(\textrm{amp})}(R_0)\delta R_0 + \frac{\mu}{N}\textrm{cos}\alpha\tilde{A}_{00}D\mathrm{H}^{(\textrm{amp})}(R_0) \delta R_0\notag \\
&= \Bigg[ \epsilon^{-1}(1-3R^2_0) + \mu\textrm{cos}\alpha  \Bigg]\delta R_0(t) \label{Eq:SL-global-syncalong-VarEQ-AMP}
\end{flalign} Hence, it gives a slightly greater Lyapunov exponent than the transverse ones 
\begin{equation}
    \Lambda_{\textrm{perturb}}^{(\textrm{amp},0)} \approx \epsilon^{-1}(1-3R^2_0) +\mu \textrm{cos}\alpha  <0 \label{Eq:SL-global-syncalong-AMP}
\end{equation} which shows that $\Lambda_{\textrm{trans}}^{(\textrm{amp},0)} \lesssim \Lambda_{\textrm{perturb}}^{(\textrm{amp},0)}$.

As for the other negative exponents, we guess that the other stable Lyapunov exponents of the amplitude DOFs are linked to the incoherent oscillators governed by the quotient dynamics in Eq.~(\ref{Eq:quotient-dynamics})
\begin{flalign}
\frac{dR_m}{dt} &=  \epsilon^{-1}(1-R^{2}_{m})R_m +\frac{\nu}{N}\tilde{A}_{m0}\textrm{cos}(s_0-s_m-\alpha)R_0  \notag \\
&+ \frac{\mu}{N}\sum_{m'=1}^{N}\tilde{A}_{mm'}R_{m'}\textrm{cos}(s_{m'}-s_m-\alpha) \label{Eq:incoh-amp-quotient}
\end{flalign} for $m=1,...,N$.

Next, we deal with the phase degrees of freedom of the Stuart-Landau oscillators ensemble. Here, we also exploit the network structure with appropriately defined coupling weights. With this approach, the governing equations for the phase DOFs read
\begin{flalign}
\frac{d \phi_i(t)}{dt} = \mathrm{F}^{(\textrm{ph})}(\phi_i(t)) + \sum_{j=1}^{2N}K_{ij}^{\textrm{(ph)}}B^{(c)}_{ij}\mathrm{H}(\phi_j(t)-\phi_i(t)) 
\end{flalign} for $i=1,...,2N$, where the uncoupled dynamics is governed by $\mathrm{F}^{(\textrm{ph})}(\phi_i)=-\sigma r_{i}^2-\frac{\mu}{N}\textrm{sin}\alpha$ and the coupling function is defined as $\mathrm{H}(x)=\textrm{sin}(x-\alpha) $. The coupling weights are defined by $K_{ij}^{\textrm{(ph)}}=\frac{\mu}{N}\frac{r_j}{r_i}$ if $i,j$ belong to the same population and $K_{ij}^{\textrm{(ph)}}=\frac{\nu}{N}\frac{r_j}{r_i}$ otherwise, provided that the amplitude variables are treated as external forcing functions. Thus, the resulting equation is the Kuramoto-type equation of Eqs.~(\ref{Eq:quotient-dynamics}). From the quotient adjacency matrix defined in Eq.~(\ref{Eq:global-quotient-adjacency-matrix}), the quotient dynamics of the synchronized and incoherent populations in phase DOFs are obtained as
\begin{flalign}
\frac{d s_0}{dt} &= -\sigma R^2_0 - \frac{\mu}{N}\textrm{sin}\alpha \notag \\
&-\frac{\mu}{N}\tilde{A}_{00}\textrm{sin}\alpha +\frac{\nu}{N}\sum_{m'=1}^{N}\frac{R_{m'}}{R_0}\tilde{A}_{0m'}\textrm{sin}(s_{m'}-s_0-\alpha) \notag \\
\frac{d s_m}{dt} &= -\sigma R^2_m-\frac{\mu}{N}\textrm{sin}\alpha+ \frac{\nu}{N}\tilde{A}_{m0} \frac{R_0}{R_m}\textrm{sin}(s_0-s_m-\alpha) \notag \\
&+\frac{\mu}{N}\sum_{m'=1}^{N}\frac{R_{m'}}{R_m}\tilde{A}_{mm'}\textrm{sin}(s_{m'}-s_m-\alpha) \notag \\
&= \tilde{\Omega}_m(t) + \nu \frac{R_0}{R_m}\textrm{sin}(s_0-s_m-\alpha) \notag \\ &+\frac{\mu}{N}\sum_{m'=1}^{N}\textrm{sin}(s_{m'}-s_m-\alpha) 
\label{Eq:SLE-phase-quotient}
\end{flalign} where $\tilde{\Omega}_m(t) = -\sigma R_{m}^{2}(t)$ for $m=1,...,N$. The quotient dynamics reveal that the phase DOFs of SL oscillator ensemble in the synchronized population also have $(N-1)$-fold degenerate transverse Lyapunov exponents with
\begin{flalign}
\Lambda_{\textrm{trans},\kappa}^{(0)} &=  -\mu \textrm{cos}\alpha -\frac{\nu}{N} \sum_{m'=1}^{N}\frac{R_{m'}}{R_0} \textrm{cos}(s_{m'}-s_0-\alpha) \notag \\ &= -\mu \textrm{cos}\alpha -\frac{\nu}{N}\tilde{\mathrm{Z}}  < 0
\end{flalign} for $\kappa=2,...,N$ where $\tilde{\mathrm{Z}}=\sum_{m'=1}^{N}\frac{R_{m'}}{R_0} \textrm{cos}(s_{m'}-s_0-\alpha)$ should be considered as an external forcing field, which follows from the transversal variational equations Eq.~(\ref{Eq:transverse-variational}): 
\begin{flalign}
\dot{\eta}_{\kappa}^{(0)} &= \Bigg[ -\frac{\mu}{N}(N-1) \textrm{cos}\alpha - \frac{\nu}{N}\tilde{\mathrm{Z}} +\frac{\mu}{N} \lambda_{\kappa}^{(0)}  \textrm{cos}\alpha  \Bigg] \eta_{\kappa}^{(0)}.
\end{flalign}Here, $\eta_{\kappa}^{(0)}(t) = \sum_{i \in C_0}u_{\kappa i}^{(0)} \delta \phi_i(t)$  for $\kappa=2,...,N$ and the deviation along the CS dynamics is $\delta \phi_i(t) = \phi_i(t)-s_0(t)$ for $i \in C_0$, and the eigenvalues $\lambda_{\kappa}^{(0)}=-1$ for all $\kappa$. In addition, the LE in the sync population coming from a perturbation along the sync-manifold has the value of $\Lambda_{\textrm{perturb}}^{(0)} =- \frac{\nu}{N}\tilde{\mathrm{Z}} <0$ and is expected to be found in the synchronized phase DOFs. 

Furthermore, we can also rationalize the eigenvalue branches of the synchronized oscillators that were already discussed in the continuum limit in Ref.~\onlinecite{Laing_SL2019} as follows. We again consider the real-valued coordinate of the SL variables in the vector form as $\bold{x}_k(t)=(a_k(t), b_k(t))^\top \in \mathbb{R}^2$ where $a_k$ and $b_k$ are defined in Eq.~(\ref{Eq:real-value-coordi}). Then, the SL oscillators evolve according to
\begin{flalign}
\frac{d}{dt}\bold{x}_i(t) &=  \bold{F}(\bold{x}_i(t)) + \sum_{j=1}^{2N} B^{(c)}_{ij}K_{ij} \bold{H}(\bold{x}_j(t))
\end{flalign} for $i=1,...,2N$ where $B^{(c)}_{ij}$ and $K_{ij}$ are defined in Eq.~(\ref{Eq:KS-global-ansatz}), the uncoupled dynamics is governed by
\begin{flalign}
\bold{F}(\bold{x}_i(t)) &= \Bigg[ \begin{pmatrix} \epsilon^{-1} & -\omega \\ \omega & \epsilon^{-1} \end{pmatrix} + \frac{\mu}{N} \begin{pmatrix} \textrm{cos}\alpha & \textrm{sin}\alpha \\ -\textrm{sin}\alpha & \textrm{cos}\alpha \end{pmatrix} \Bigg] \bold{x}_{i}(t) \notag \\ &- \frac{\epsilon^{-1}}{2} \begin{pmatrix} 1 & -\epsilon \sigma \\ \epsilon \sigma & 1 \end{pmatrix}|\bold{x}_{i}(t)|^2 \bold{x}_{i}(t)
\label{Eq:uncoupled}
\end{flalign} and the coupling function is written as
\begin{flalign}
\bold{H}(\bold{x}_i(t)) = \begin{pmatrix} \textrm{cos}\alpha & \textrm{sin}\alpha \\ -\textrm{sin}\alpha & \textrm{cos}\alpha \end{pmatrix}  \bold{x}_{i}(t)
\label{Eq:coupling function}
\end{flalign} for $i=1,...,2N$. If we also regard the chimera state as a CS pattern dynamics: $\bold{x}_i(t) = \bold{s}_0$ (sync.) and $\bold{x}_{i+N}(t) = \bold{s}_m(t)$ (incoh.) for $i=m=1,...,N$, then the variational equations transversal to the synchronized cluster $C_0$ in the cluster-based coordinates are given by
\begin{flalign}
\dot{\bf{\eta}}_{\kappa}^{(0)}=&  \bigg[ D\bold{F}(\bold{s}_0)+ \frac{\mu}{N} \lambda_{\kappa}^{(0)}D\bold{H}(\bold{s}_0) \bigg] \bf{\eta}_{\kappa}^{(0)}
\end{flalign} for $\kappa=2,...,N$ where the Jacobians of the dynamical functions in Eqs.~(\ref{Eq:uncoupled}-\ref{Eq:coupling function}) read 
\begin{flalign}
D\bold{F}(\bold{s}_0) &= \begin{pmatrix} \epsilon^{-1} & -\omega \\ \omega & \epsilon^{-1} \end{pmatrix} + \frac{\mu}{N} \begin{pmatrix} \textrm{cos}\alpha & \textrm{sin}\alpha \\ -\textrm{sin}\alpha & \textrm{cos}\alpha \end{pmatrix} \notag \\
&- \frac{\epsilon^{-1}}{2} \begin{pmatrix} 1 & -\epsilon \sigma \\ \epsilon \sigma & 1 \end{pmatrix} \begin{pmatrix} 3 s_{0_1}^2+s_{0_2}^2 & 2 s_{0_1}s_{0_2} \\  2s_{0_1}s_{0_2} &3 s_{0_2}^2+s_{0_1}^2\end{pmatrix}
\end{flalign} and 
\begin{flalign}
D\bold{H}(\bold{s}_0) =  \begin{pmatrix} \textrm{cos}\alpha & \textrm{sin}\alpha \\ -\textrm{sin}\alpha & \textrm{cos}\alpha \end{pmatrix} .
\end{flalign} Since for the synchronized SL oscillators we have $r_k e^{i\phi_k} = e^{i \phi_0} = \frac{1}{\sqrt{2}}(a_0+i b_0)$ for $k=1,...,N$, we can rewrite the transversal variational equations in the following form
\begin{flalign}
\dot{\bf{\eta}}_{\kappa}^{(0)} &= \Bigg[ \begin{pmatrix} \epsilon^{-1} & -\omega \\ \omega & \epsilon^{-1} \end{pmatrix}+ \frac{\mu}{N}(1+\lambda_{\kappa}^{(0)})  \begin{pmatrix} \textrm{cos}\alpha & \textrm{sin}\alpha \\ -\textrm{sin}\alpha & \textrm{cos}\alpha \end{pmatrix}   \notag \\ & - \frac{\epsilon^{-1}}{2} \begin{pmatrix} 1 & -\epsilon \sigma \\ \epsilon \sigma & 1 \end{pmatrix} \begin{pmatrix} 2+4\textrm{cos}^2\phi_0 &  4\textrm{cos}\phi_0 \textrm{sin}\phi_0 \\ 4\textrm{cos}\phi_0 \textrm{sin}\phi_0 & 2+4\textrm{sin}^2\phi_0 \end{pmatrix} \Bigg] \bf{\eta}_{\kappa}^{(0)} \notag \\
&=\bold{J}^{(0)}_{\textrm{trans}} \bf{\eta}_{\kappa}^{(0)}
\end{flalign} for all the directions transverse to the sync-manifold. Notice that the eigenvalues of the adjacency matrix $\lambda_{\kappa}^{(0)} =-1$ for all $\kappa$. If we consider $\phi_0$ as an external forcing function, then the eigenvalues of the matrix $\bold{J}^{(0)}_{\textrm{trans}}$ are
\begin{flalign}
\Lambda_1 &= -\frac{1+\sqrt{1-\epsilon^{2}(3\sigma^2-4\sigma \omega +\omega^2)  }}{\epsilon}  \notag \\
\Lambda_2 &= \frac{-1+\sqrt{1-\epsilon^{2}(3\sigma^2-4\sigma \omega +\omega^2)  }}{\epsilon} 
\end{flalign} which gives $\Lambda_1 \sim -2\epsilon^{-1} $ corresponding to the amplitude DOF branch and $\Lambda_2 \lesssim 0$ corresponding to the phase DOF branch for the synchronized oscillators. This result and our previous analysis strongly suggest that the negative branch indeed arises from the amplitude DOFs and the near-zero branch comes from the phase DOFs including slow and stable Lyapunov exponents, and both render the Poisson chimeras attracting. 

\section{Concurrent dynamical and topological variations: Stuart-Landau oscillators on nonlocal intra-population topology\label{append:concurrent}}

Here, both the topological and dynamical variations are introduced simultaneously. Thus, we consider Stuart-Landau amplitude oscillators in the nonlocal intra-population network topology, and focus on weak coupling with ($\epsilon=0.01$). Starting from $\textbf{PIC}$, we observe chimera states that are similar to those in Sec.~\ref{subsec:nonlocal}. Hence, the Poisson chimeras with the parameters $A=0.2$ and $A=0.35$ follow the similar incoherent dynamics as in Fig.~\ref{Fig:Nonlocal-KS-order-parameter}.

The Lyapunov analysis for the nonlocal Stuart-Landau oscillators obviously results in the properties dictated by the given nonlocal topology of the network. From the same method discussed in the previous sections,
we obtain the two different values of the degenerate transverse Lyapunov exponents in the synchronized group of phase DOFs
\begin{flalign}
\Lambda_{\textrm{trans},\kappa}^{(0)}&=- \frac{\mu}{N}(N-2)\textrm{cos}\alpha+\frac{\mu}{N}\lambda_{\kappa}^{(0)} \textrm{cos}\alpha-\frac{\nu}{N}\tilde{\mathrm{Z}} \notag \\  &= 
\begin{dcases}
    -\frac{\mu}{N}(N-2)\textrm{cos}\alpha -\frac{\nu}{N}\tilde{\mathrm{Z}} , & \kappa =  2,...,N/2+1 \\
     - \mu\textrm{cos}\alpha -\frac{\nu}{N}\tilde{\mathrm{Z}}, & \kappa=N/2+2,...,N
  \end{dcases}
  \notag
\end{flalign} where  $\lambda_{\kappa}^{(0)}=0$ for $\kappa=2,...,N/2+1$ and  $\lambda_{\kappa}^{(0)}=-2$ for $\kappa=N/2+2,...,N$. Also, the negative LE corresponding to the sync-manifold perturbation is given as $\Lambda_{\textrm{perturb}}^{(0)} =- \frac{\nu}{N}\tilde{\mathrm{Z}} <0$, strongly depending on the collective behavior of the incoherent oscillators. Finally, in the incoherent population, we obtain the same $N/2$ pairs of the two nearly-degenerate exponents that result from the discrete symmetries of the phase governing equations of the Stuart-Landau oscillators. Therefore, the Poisson chimera trajectories of this system are also attracting more strongly than other cases.

\begin{figure}[t!]
\includegraphics[width=1.0\linewidth]{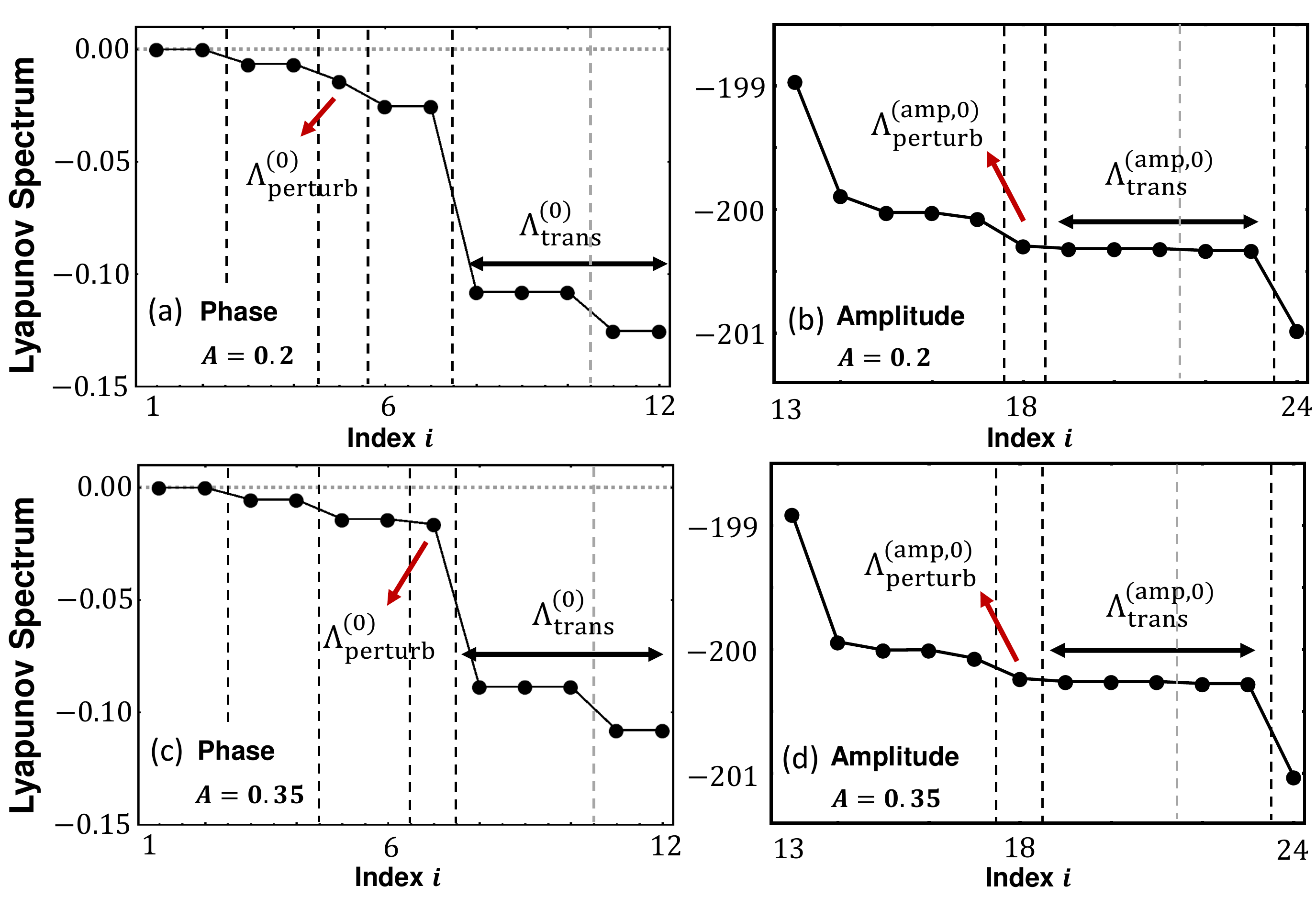}
\caption{Full Lyapunov spectra of the Stuart-Landau oscillators with nonlocal intra-population topology for (a) the phase degrees of freedom  and (b) the amplitude degrees of freedom. The parameter set used here is $N=6$, $A=0.2$ and $\epsilon=0.01$. (c-d) The breathing chimera states with $A=0.35$. Note that the Lyapunov exponents in (a) follow the same behavior as the stationary chimera state of the phase-only system (compare Fig.~\ref{Fig:LE-nonlocal-phase}).} 
\label{Fig:LE-nonlocal-SLE}
\end{figure}

Regarding the amplitude DOFs, the $N-1$ transverse Lyapunov exponents also show the two different values of the degenerate exponents approximated as \begin{flalign}
\Lambda_{\textrm{trans},\kappa}^{(\textrm{amp},0)} &= \epsilon^{-1}(1-3R^2_0)+\frac{\mu}{N}(1+\lambda_{\kappa}^{(0)})\textrm{cos}\alpha \\ &= 
\begin{dcases}
    \epsilon^{-1}(1-3R^2_0)+\frac{\mu}{N}\textrm{cos}\alpha , & \kappa =  2,...,N/2+1 \\
     \epsilon^{-1}(1-3R^2_0)-\frac{\mu}{N}\textrm{cos}\alpha, & \kappa=N/2+2,...,N
  \end{dcases}
  \notag
\end{flalign}
since for the nonlocal network $\lambda_{\kappa}^{(0)}=0$ for $\kappa=2,...,N/2+1$ and $\lambda_{\kappa}^{(0)}=-2$ for $\kappa=N/2+2,...,N$ (distinguished by the gray dashed line in Fig.~\ref{Fig:LE-nonlocal-SLE} (b,d)). Then, we expect to find the sync-manifold perturbation exponent of the amplitude DOFs, $\Lambda_{\textrm{perturb}}^{(\textrm{amp},0)} \approx  \epsilon^{-1}(1-3R^2_0) +\frac{\mu}{N}(N-1)\textrm{cos}\alpha  $ which is slightly greater than the transverse exponents. As for the other exponents, we only know that they arise from the incoherent governing equations.

Judging from the above observation, we conclude that the Poisson chimera states are definitely attracting. 
A comparison with the systems that have only one `perturbation' compared to the globally coupled phase oscillators, i.e. either the non-local coupling topology or the amplitude DOF, suggests that the attraction rate of the phase DOF is mainly determined by the non-local network topology.

\vskip 1cm 

\bibliography{aipchaos}


\end{document}